\documentclass[journal]{IEEEtran}

\usepackage[english]{babel}
\usepackage{color,soul}
\usepackage{xcolor}

\usepackage{graphicx} 
\usepackage{url} 
\usepackage[small]{subfigure}
\usepackage{amsmath}
\usepackage{bm}
\usepackage{multirow, array}

\usepackage{makecell}
\usepackage{balance}
\usepackage{tabu}
\usepackage{algorithm}
\usepackage{algorithmic}

\usepackage{enumitem}
\newcommand{\rev}[1]{\textcolor{black}{#1}}

\usepackage{pifont}

\begin{document}

\title{FALCON: Fast and Accurate Multipath Scheduling using Offline and Online Learning}

\author{Hongjia Wu, \"{O}zg\"{u} Alay, Anna Brunstrom, Giuseppe Caso, Simone Ferlin}

\maketitle

\begin{abstract}
Multipath transport protocols enable the concurrent use of different network paths, benefiting a fast and reliable data transmission. The scheduler of a  multipath transport protocol determines how to distribute data packets over different paths. Existing multipath schedulers either conform to  predefined policies or to  online trained policies. The adoption of millimeter wave (mmWave) paths in 5th Generation (5G) networks and Wireless Local Area Networks (WLANs) introduces time-varying network conditions, \rev{under which the existing schedulers struggle to achieve fast and accurate adaptation.} In this paper, we propose FALCON, a learning-based multipath scheduler that can adapt fast  and accurately to time-varying network conditions. \rev{FALCON builds on the idea of \emph{meta-learning} where offline learning is used to create  a  set  of  meta-models  that  represent coarse-grained network conditions, and online learning is used to bootstrap a specific model for the current fine-grained network conditions towards deriving the  scheduling  policy to deal with such conditions.} 
Using trace-driven emulation experiments, we demonstrate FALCON outperforms the best state-of-the-art scheduler by up to 19.3\% and 23.6\% in static and mobile networks, respectively. \rev{Furthermore, we show FALCON is quite flexible to work with different types of applications such as bulk transfer and web services. Moreover, we observe FALCON has a much faster adaptation time compared to all the other learning-based schedulers, reaching almost an 8-fold speedup compared to the best of them.} Finally, we have validated the emulation results in real-world settings illustrating that FALCON adapts well to the dynamicity of real networks, consistently outperforming all other schedulers.

\end{abstract}

\section{Introduction}
\label{intro}

The 5th Generation of mobile communications (5G) raises the expectations towards three key performance aspects: very high data rates, ultra-reliable and low-latency communications, and massive connectivity. To accommodate these requirements, the concurrent use of multiple Radio Access Technologies (RATs), i.e., \emph{multi-connectivity}, is one of the key solutions highlighted in 5G systems~\cite{andrews2014will}. Among several 5G multi-connectivity schemes~\cite{suer2019multi}, multipath transport protocols, such as multipath Transmission Control Protocol (MPTCP)~\cite{rfc8684} and multipath QUIC (MPQUIC)~\cite{DeConinckQUIC}, have recently gained significant attention. In particular, this is due to the Technical Specification (TS) 23.501 (Release 16) by 3$^{\mathsf{rd}}$ Generation Partnership Project (3GPP)~\cite{TS23501}, where it is discussed how 5G systems can take advantage of multipath transport protocols to support the Access Traffic Steering, Switching and Splitting (ATSSS) architecture, ultimately enabling multi-connectivity between 3GPP access, such as Long Term Evolution (LTE) and 5G New Radio (NR), and non-3GPP Wireless Local Area Networks (WLAN), such as WiFi. 

Among the functionalities of multipath transport protocols, the multipath scheduler plays a key role since it regulates the distribution of data packets over different available paths (i.e., the available RATs), ultimately impacting the achievable performance in terms of experienced throughput, latency, and connection reliability. The design of a high-performing multipath scheduler is a challenging problem, especially under high time-varying network conditions, e.g., in the case of millimeter wave (mmWave) paths with high propagation losses and sensitivity to blockage~\cite{WuCommag21}. To operate well in such challenging conditions, a multipath scheduler should be able to meet two main targets: a) \textit{fast adaptation}, i.e., adapt its scheduling policy quickly to the network conditions, and b) \textit{accurate adaptation}, i.e., the scheduling policy should capture the network conditions accurately.

Existing multipath schedulers are either based on predefined rules (e.g., using the path with minimum Round Trip Time (RTT)) or on Machine Learning (ML) schemes (e.g., using a Reinforcement Learning (RL) algorithm to select the best path to use under some specific network conditions). Schedulers based on predefined rules define a priori, rules based on the network conditions that they will adapt to (cf. Section \ref{schedulers}). As these schedulers do not need to \emph{learn} the scheduling policy to use, their adaptation time is negligible meeting the fast adaptation target. However, the predefined rules often result in a coarse-grained scheduling policy that might not adapt well to the current network conditions, particularly when such conditions vary rapidly. Thus, schedulers based on predefined rules have difficulty in meeting the accurate adaptation target.

Schedulers based on ML use, in particular, \emph{online learning} approaches, observe the current network conditions and adapt to them by deriving a corresponding scheduling policy (cf. Sections \ref{schedulers} and \ref{learning:networking}). Compared to schedulers based on predefined rules, they require extra time for \textit{learning} the policy, thus resulting in slower but possibly more accurate adaptation to network conditions.
In fact, a trade-off exists for these schedulers in terms of fast vs. accurate adaptation. 
On the one hand, if the scheduler (i.e., the learning agent) employs a complex learning architecture, e.g., a deep neural network~\cite{zhang2019reles,rosello2019multi}, it may converge to an accurate policy but this may require more time, thus inhibiting fast adaptation. On the other hand, if the scheduler employs a simple learning scheme, e.g., a lightweight RL algorithm~\cite{wu2020peekaboo,wumultipath}, it may converge faster at the cost of accuracy.

To address the above challenges faced by online learning schedulers, we argue that scheduling operations may benefit from further training based on \emph{offline learning}. Indeed, a scheduler may use previous experience on already faced network conditions for deriving proper model(s) for newly encountered conditions; then, such model(s) can be exploited by the online learning algorithm for obtaining a fast and accurate scheduling policy. This idea is further clarified and justified throughout Sections \ref{foundation} and \ref{problem}. Within the above context, this paper proposes FALCON, a ML-based multipath scheduler that combines online and offline learning. 
FALCON builds on the idea of \emph{meta-learning}~\cite{finn2017model, nichol2018first}, where a meta-model is set up via offline learning and fine-tuned via online learning. The online learning experience also feeds back to the offline learning function to form a closed loop for continuously updating the meta-model. The contributions of our work can be summarized as follows:

\begin{itemize}
    \item \rev{We present the necessity for a multipath scheduler to be able to adapt fast and accurately to varying network conditions and show that existing multipath schedulers have difficulty to meet this objective;}
	\item We design FALCON, an ML-based multipath scheduler that combines the benefits of offline and online learning \rev{for deriving trained multipath scheduling policies with a reduced amount of input data. To the best of our knowledge, our work is the first systematic study on multipath scheduling that optimizes both adaptation speed and accuracy to time-varying network conditions;}
	\item We implement the protocol aspects of FALCON in MPQUIC using \texttt{quic-go} and the learning aspect of FALCON using \texttt{keras-rl}.
	All software components of FALCON are provided as open-source to the community.\footnote{Upon acceptance of the paper.}
	\item Using trace-driven emulations, \rev{we demonstrate fast and accurate adaptation and thus the superior performance of FALCON for applications of bulk transfer and web service with multi-streaming support compared to the state-of-the-art multipath schedulers.} 
	\item \rev{We validate the emulation results in real-world settings and show that FALCON outperforms all other schedulers in realistic network conditions.}
	\end{itemize}

The rest of this paper is organized as follows. We first summarize the foundations and related work of our work in Section~\ref{foundation}. We then specify the research problem and provide an overview of FALCON in Section~\ref{problem}. We next detail the design of FALCON in Section~\ref{Design:O2}. We present the experimental setup in Section~\ref{setup} and evaluate the  performance of FALCON via emulations in Section~\ref{emulation} and real-world experiments in Section~\ref{realworld}. We finally conclude our work in Section~\ref{conclusion}.

\section{Foundations and Related Work}
\label{foundation}
In this section, we summarize foundations and related work of FALCON, including aspects related to multipath transport (Section~\ref{multipath_transport}), multipath scheduling (Section~\ref{schedulers}), and learning in networking scenarios (Section~\ref{learning:networking}).  

\subsection{Multipath Transport Protocol} 
\label{multipath_transport}
Multipath transport protocols are designed to achieve higher throughput and resilience compared to their single-path counterparts, since they can leverage several paths simultaneously and support seamless failover. In particular, two multipath protocols have wide support from both standardization and research communities: MPTCP and MPQUIC.

MPTCP~\cite{rfc8684} is the multipath extension of TCP and has the goal of being transparent to both higher and lower protocol layers. Its design and operation are influenced by the proliferation of middleboxes, meddling in end-to-end TCP connections, and preventing TCP extensions as well as the deployment of new transport protocols. Adopting several successful features of TCP, QUIC became recently an attractive alternative, as it integrates Transport Layer Security (TLS) 
and improves latency at the connection start. Differently from TCP, QUIC encrypts most of the protocol headers and all payloads to prevent interference from middleboxes. Motivated by the success of MPTCP, there are already some MPQUIC implementations proposed as multipath extensions of QUIC \cite{DeConinckQUIC,LiuMPQUIC2021}. We leverage MPQUIC to perform the analysis of multipath schedulers in this paper, as we believe it will play a key role in determining the multi-connectivity performance in 5G.

\subsection{Multipath Scheduling}
\label{schedulers}
The multipath scheduler is in charge of distributing packets over the available paths. In the following, we describe two categories of multipath schedulers: Based on predefined rules and based on ML schemes.

\vspace{1mm}
\noindent\textbf{\textit{Schedulers based on predefined rules}}: Traditional multipath schedulers follow predefined rules that do not change over time. 
For example, a Round Robin (RR) scheduler cyclically sends packets over each path, as long as there is space in the congestion window (CWND) of the paths. RR may perform reasonably well when the available paths have similar characteristics (i.e., paths are \emph{homogeneous}). However, since it does not consider the characteristics of the individual paths it is unable to prevent out-of-order packet arrival at the receiver, which is detrimental to multipath transport performance. The minimum RTT (minRTT) scheduler has shown that considering and exploiting path characteristics, e.g., by sending packets on the path with available CWND and lowest RTT, allows achieving higher throughput~\cite{ferlin2016blest}. Indeed, minRTT is the default scheduler in both MPTCP and MPQUIC.

Other schedulers based on predefined rules have been proposed over the years. Blocking Estimation (BLEST)~\cite{ferlin2016blest} and Earliest Completion First (ECF)~\cite{lim2017ecf} try to provide both high throughput and low latency. Assuming two available paths, when both paths have CWND availability, BLEST and ECF behave like minRTT, i.e., they select the path with the lowest RTT. When the path with the lowest RTT has no CWND availability, BLEST and ECF use different mechanisms to decide whether it is better to send packets on the path with the highest RTT or wait for the path with the lowest RTT to become available again. Addressing specific use cases and applications, the works in \cite{frommgen2016remp,lee2018raven, guo2017accelerating} apply an adaptive packet duplication mechanism to guarantee robustness, which proves to be effective when extra data usage and battery consumption are not limiting factors. The work in \cite{shi2018} proposes the Slide Together Multipath Scheduler (STMS) to reduce out-of-order packet arrivals and, thus, the receiver buffer problem. \cite{dong2019loss} proposes a loss-aware scheduler targeting networks with more than 20\% loss rates.~\cite{hurtig2019} proposes the Short Transfer Time First (STTF) scheduler, targeting low latency for short transfers and considering TCP specific aspects such as the TCP Small Queues (TSQ). Lastly,~\cite{saha2019musher} proposes a multipath scheduler for MPTCP that targets IEEE 802.11 ad/ac WLANs. 

\vspace{1mm}
\noindent\textbf{\textit{Schedulers based on ML}}:
Nevertheless, confronted with the complexity of the network conditions, it is difficult for schedulers based on predefined rules to guarantee the accuracy for various environment characteristics. Multipath scheduling can be also thought of as a decision-making problem thus naturally fitting into scenarios that RL schemes aim to solve, including multi-armed bandit problems (MAB) and Markov decision processes (MDP). For this reason, there is an emerging interest in developing ML-based multipath schedulers. Adopting the MAB framework,~\cite{wu2020peekaboo} combines the Linear Upper Confidence Bound (LinUCB) algorithm and a stochastic adjustment to design a multipath scheduler in MPQUIC, namely Peekaboo, that shows improved performance in dynamic heterogeneous networks compared to schedulers based on predefined rules.  Then,~\cite{wumultipath} proposes Modified-Peekaboo (M-Peekaboo) by extending the learning scheme of Peekaboo for path selection, aiming at extending the applicability range towards 5G mmWave networks. Framing the scheduling problem as an MDP,~\cite{zhang2019reles} exploits the Deep Q-Network (DQN) architecture to design a multipath scheduler in MPTCP, namely Reles, which shows performance gains over minRTT.  Similarly,~\cite{rosello2019multi} also designs a multipath scheduler using DQN in MPQUIC, resulting in no clear performance gain over minRTT. 

\subsection{Learning Concepts in Networking}
\label{learning:networking}
As mentioned in Section \ref{intro}, our proposed scheduler, FALCON, belongs to the category of schedulers based on ML. However, as also clarified later, we aim at not only leveraging previous scheduling approaches, all based on online learning but also to include offline learning, to improve the overall performance. Hence, in this section, we provide an overview of both offline and online learning approaches currently considered and adopted in networking applications more generally, thus, not limited to multipath scheduling. Then, we also provide a high-level description of meta-learning, which is the actual framework used in FALCON for leveraging offline and online learning functionalities.

\vspace{1mm}
\noindent\textbf{\textit{Offline learning:}} This paradigm assumes that, in order to derive a model of and/or a policy for a generic environment, an ML algorithm uses environment characteristics, i.e., data, collected well-ahead, before the derived model is meant to be used. In the following, we refer to pre-collected data as \emph{offline data}. The learning outcome, e.g., the policy to be used by a network protocol, is not modified once derived on offline data. In other words, there is no retraining. Therefore, the assumption is that offline data includes a complete enough set of environment characteristics that could be experienced when the model/policy is actually used. 
To mention a few, offline learning is used to derive offline data-based policies for congestion control using optimization approach~\cite{winstein2013tcp}, Adaptive Bit Rate (ABR) streaming using DQN~\cite{akhtar2018oboe} or Asynchronous Advantage Actor Critic (A3C)~\cite{mao2017neural}, and device resource management using DQN~\cite{mao2019learning} or Support Vector Machine (SVM)~\cite{ren2018proteus}. To the best of our knowledge, offline learning is not currently used for multipath scheduling. 

\vspace{1mm}
\noindent\textbf{\textit{Online learning:}} This paradigm assumes that to derive a model and/or policy, an ML algorithm uses data that is collected while the model/policy is being derived and used. In the following, we refer to run-time collected data as \emph{online data}. Differently from the offline learning paradigm, the learning outcome is thus modified and adapted at run-time, exploiting newly encountered environment characteristics, i.e., new online data. This is commonly performed via two main approaches, i.e., with or without the use of an \emph{abandoning mechanism}. In the first approach, the model/policy is abandoned when either a significant change in the environment characteristics is detected via so-called \emph{change point detection}~\cite{wu2020peekaboo,padmanabha2018mitigating}, or a predefined timer expires~\cite{dong2018pcc,jiang2017pytheas, gilad2020mpcc}. 
Peekaboo~\cite{wu2020peekaboo} and M-Peekaboo~\cite{WuCommag21} are relevant examples of schedulers that use an online learning approach with the change point detection. In the second approach, the online learning algorithm does not apply the abandoning mechanism, i.e., the model/policy is continuously updated since the algorithm is continuously fed with online data~\cite{li2019smartcc}. Hence, in this case, there is no abrupt model/policy abandoning, which may cause a slower reaction to sudden changes in the environment characteristics. Examples of multipath schedulers that use online learning with no abandoning mechanisms are~\cite{rosello2019multi,zhang2019reles}.     
It is worth mentioning that the aforementioned online learning approaches may face the well-known catastrophic forgetting problem~\cite{kemker2018measuring}. Indeed, with or without abandoning mechanisms, the continuous feed of online data may result in the derivation of new models/policies; old models/policies that resulted to be optimal for specific environment characteristics may thus be discarded, and they need to be re-discovered if the same environment characteristics reappear.  
As a remedy, for example, ~\cite{huang2020quality} tries to apply lifelong learning for video streaming to alleviate the catastrophic forgetting problem. 

\vspace{1mm}  
\noindent\textbf{\textit{Meta-learning:}}  The meta-learning paradigm, also known as “learning to learn”~\cite{thrun2012learning}, combines online and offline learning. The goal of meta-learning is to derive (offline) a so-called meta-model for the set of learning tasks an ML algorithm needs to solve. The meta-model is built so that it can be rapidly adapted (online) to any new learning task that may be encountered, exploiting just a few experiences from the new task. 
The works in~\cite{finn2017model, nichol2018first} validate a meta-learning framework that can be used in several learning tasks, e.g., it can be applied to both supervised ML (regression and classification) and RL scenarios. Other works propose meta-learning for more specific scenarios, i.e., the update rule and selective copy of weights of deep networks~\cite{andrychowicz2016learning, ravi2016optimization, schmidhuber2004optimal} and recurrent networks~\cite{santoro2016meta, wang2016learning, munkhdalai2017meta}. 
In this paper, we design FALCON based on meta-learning paradigm to obtain fast and accurate scheduling policies.

\section{Problem statement and solution overview}
\label{problem}
In this section, we explain the research problem (Section \ref{problem_statement}) and provide an overview of our solution (Section \ref{solution_overview}). 

\subsection{Problem Statement}
\label{problem_statement}
The network conditions faced by a multipath scheduler vary in time, due to network congestion, users' mobility, dynamic characteristics of the wireless channels, etc. The recent use of mmWave spectrum in cellular networks and WLANs further increases this variability~\cite{narayanan2020first, zhou2018ieee}. 
Hence, as highlighted in Section \ref{intro}, a multipath scheduler should be able to adapt fast and accurately to challenging time-varying network conditions. Upon detection of a change of the network condition by the scheduler, adapting fast indicates that the adaptation time for realizing the adapted policy should be as small as possible; adapting accurately indicates that the adapted policy should match the current network condition as much as possible. 
This is, however, a difficult task, and \textit{the design of a multipath scheduler that can adapt fast and accurately to time-varying network conditions} is an open research problem. 

In the following, we clarify the limitations of existing schedulers (based on either predefined rules or learning paradigms) in meeting the above objective. Then, we also analyze the limitations that schedulers based on a pure offline learning approach would face. This analysis serves for further motivating our approach in designing FALCON, summarized in Section \ref{solution_overview} and detailed in Section \ref{Design:O2}, where we combine the benefits of offline and online learning approaches.

\textit{Schedulers based on predefined rules} can adapt fast but not accurately to time-varying network conditions. This is due to the inherent limitation caused by predefining the rule to follow for scheduling packets over the available paths. Indeed, the rule is usually rather simple and coarse-grained (e.g., select the path with minimum average RTT), thus failing to adapt accurately to the complex dynamics of the network conditions. 
	
\textit{Schedulers based on online learning} can ensure the derivation of an accurate scheduling policy. In general, however, the need for learning the network conditions online makes the adaptation slower compared to schedulers based on predefined rules. In order to speed up adaptation, schedulers based on online learning can sacrifice accuracy, thus exploiting a limited amount of data (observed network conditions) and a simple learning architecture for deriving a policy. In the following, we refer to these schedulers as \emph{Type-I} online learning based schedulers. As empirically shown in~\cite{wumultipath}, state-of-the-art Type-I schedulers still face challenges in satisfying the requirements in terms of adaptation time of modern networks, e.g., 5G mmWave. If accurate adaptation is preferred over fast adaptation, the online scheduler can exploit a larger amount of data and a more complex learning model. In the following, we refer to these schedulers as \emph{Type-II} online learning based schedulers.  
	
\textit{Schedulers based on offline learning} may intuitively seem like a reasonable approach for achieving both fast and accurate adaptation. An offline learning-based scheduler may adapt fast because it is pre-trained. Moreover, such a scheduler might achieve accurate adaptation if trained on all the possibly encountered network conditions. However, this assumption is rather unrealistic for two main reasons: (1) Collecting all possible network conditions (past and future) is nearly impossible~\cite{shi2021adapting, rotman2020online}; (2) Even if all combinations of network conditions could be found, it is difficult to accurately label each of them mathematically. Hence, several combinations of network conditions may be involved in the pre-training, and the obtained model would still have a coarse-grained match with the fine-grained network conditions. 

\subsection{Solution Overview}
\label{solution_overview}
In this work, we design and implement a scheduler based on learning. We make this choice since, compared to schedulers based on predefined rules, schedulers based on learning have the ability to learn from the encountered network conditions and adapt to their variability over time. 

\begin{figure}[t]
	\centering
	{\includegraphics[width=0.9\columnwidth]{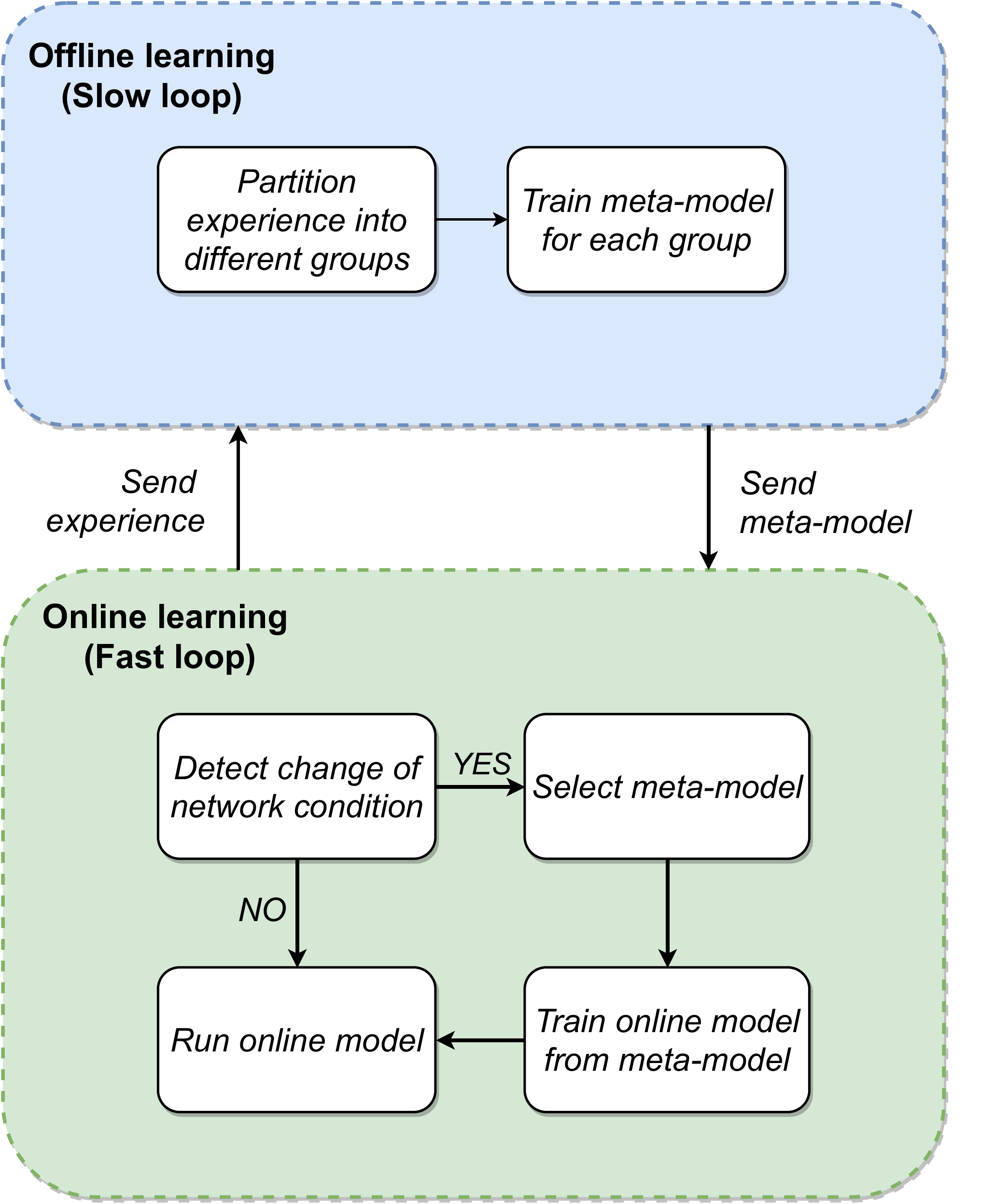}}
	\caption{An illustration of the FALCON architecture.}
	\label{fig_algorithminfo}
\end{figure}

To solve the research problem and overcome the limitations of online-only and offline-only learning-based schedulers outlined above, we propose FALCON, a multipath scheduler that combines offline learning and online learning. The key idea in FALCON is to use the meta-learning framework as offline learning to create a set of meta-models that represent the network conditions. Then, the set of meta-models is used by an online learning algorithm to bootstrap a specific model for the current network conditions and derive the scheduling policy to deal with such conditions. The meta-models are created so that they can converge to any specific model with only a small amount of online data.

On the one hand, the set of meta-models is a common root for the specific models. It is a sort of global view and thus, in contrast to the specific models, it is not very sensitive to the variability of network conditions. Hence, it can be updated at a relatively slow rate. On the other hand, online learning performs a fine-tuning of the meta-models and eventually obtains the specific model that suits the current network conditions. Hence, online learning operations are performed at a faster rate so to cater to the changes in network conditions. In other words, the model update is split into low-frequency and high-frequency updates.  

Compared to the offline-only learning approach, FALCON has the ability to adapt to the current environment without labeling the current network conditions specifically to avoid the issues of both dealing with unseen network conditions and matching a coarse-grained model to a fine-grained network condition. Compared to Type-I online learning based schedulers, FALCON efficiently uses more data and a refined learning architecture, thus achieving a higher accuracy without sacrificing fast adaptation. Compared to Type-II online learning based schedulers, the loop of creation and refinement of meta-models allows achieving faster adaptation without sacrificing accuracy. Figure~\ref{fig_algorithminfo} illustrates the FALCON architecture whose main functions can be summarized as follows:

\vspace{1mm}
\noindent\textbf{\textit{Offline Learning:}} Based on the experiences from the online learning module, the offline learning module partitions the experiences into different groups, based on the network conditions. For each group of experiences, the offline learning module performs meta-learning and derives a meta-model. The offline learning is set up tackling the changes of meta-model in real-world scenarios. The shared knowledge stored in the meta-model does not change in a very fast manner but rather an extremely gradual manner to slightly adjust with the internal representation in the real-world, thus updating in a very slow frequency. 
  
\noindent\textbf{\textit{Online Learning:}} The online learning module continuously monitors the change of network conditions. Depending on the outcome of the change detection, the online learning module may choose to deploy the current model (no changes are detected) or perform the model retraining (a change is detected). In the second case, the online learning module performs a training from a meta-model selected based on the group where the current network condition belongs. Thus, the online learning loop updates in a fast frequency. 
 
\noindent\textbf{\textit{Information Exchange:}}
Online and offline modules cooperate and exchange experiences and meta-models in a recursive loop, as also shown in Figure~\ref{fig_algorithminfo}.

\section{FALCON Design}
\label{Design:O2}
In this section, we describe FALCON's algorithm by presenting its pseudo-code and the learning strategies adopted in the offline and online learning modules (Section~\ref{overall}). Then, we further specify the learning elements needed for setting up FALCON operations (Section~\ref{elements}).

\begin{algorithm} [t]
	\caption{~FALCON pseudo-code.}\label{algorithm_O2}
	\begin{algorithmic}[1]
		\REQUIRE~~\\
		1$)$ $T_{\mathsf{upd}}$: Update interval of meta-models;\\
		  2$)$ $\bm{R}_{\mathsf{S}}$: Set of network condition ranges for meta-models;\\
	    \hskip\dimexpr-1.7\algorithmicindent $\textbf{Offline learning module:} $\\
	    \WHILE {True}
	    \STATE {$\bm{Exp} = \text{CollectOnlineExperience}();$}\\
		\STATE {$\bm{\Theta}_{\mathsf{S}} = \text{MetaLearn}(\bm{Exp}, \bm{R}_{\mathsf{S}});$}\\
		\STATE {$\text{Wait}(T_{\mathsf{upd}});$}\\
		\ENDWHILE\\
		\hskip\dimexpr-1.75\algorithmicindent $\textbf{Online learning module:} $\\
		\WHILE {True}
		\STATE {$\bm{\Theta}_{\mathsf{S}} = \text{CollectOfflineMetaModels}();$}\\
		\STATE {$\bm{Exp} = \text{Execute}(\textbf{\text{CurrentPolicy}});$}\\
		\STATE {$\text{DetectedChange} = \text{ChangeDetect}();$}\\
		\IF{ $\text{DetectedChange}$}
		\STATE {$\bm{\Theta} = \text{SelectMetaModel}(\bm{R}_{\mathsf{S}}, \bm{\Theta}_{\mathsf{S}});$\\}
		\STATE {$\textbf{\text{NewPolicy}} = \text{FewShotLearn}(\bm{\Theta});$\\}
		\STATE {$\textbf{\text{CurrentPolicy}} \leftarrow\textbf{\text{NewPolicy}};$}\\
		\ENDIF \\
		\ENDWHILE
	\end{algorithmic}
\end{algorithm}

\subsection{Algorithm}
\label{overall}
Algorithm~\ref{algorithm_O2} reports FALCON pseudo-code. As also illustrated in Figure \ref{fig_algorithminfo}, FALCON leverages both offline and online learning via dedicated modules that exchange current experience and meta-models in a recursive loop. In the following, we provide more details on both modules. 

\vspace{1mm}
\noindent \textbf{\textit{Offline learning module:}} This module derives a set of meta-models that represent, on a high-level, the network conditions under which FALCON operates. The meta-models enable the online learning module to timely derive an accurate scheduling policy for the current network conditions. To do so, FALCON leverages the concept of meta-learning, where the main idea is to find a 
meta-model, denoted $\bm{\Theta}$, for solving a generic learning \emph{task}. 
Meta-model $\bm{\Theta}$ represents the common starting point from which a number of refined models, that map onto more specific learning tasks, can be derived. 
For example, a meta-model is a high-level knowledge that a RL agent may have on how to navigate mazes (i.e., a generic task). Then, when the agent is deployed in a maze with specific characteristics (i.e., a specific task), it can exploit the high-level knowledge so to quickly learn how to navigate that specific maze \cite{finn2017model}. Indeed, $\bm{\Theta}$ is created so that the refined models that match specific tasks can be derived in a few gradient steps. The requirement of $\bm{\Theta}$ is that starting from $\bm{\Theta}$, the online model can converge within several online gradient steps to match the presented network condition. In other words, $\bm{\Theta}$ guarantees that a few-shot learning~\cite{sung2018learning, sun2019meta} is sufficient for finding the refined models. Considering that one online model may have several convergence points within the parameter space subject to the common machine learning paradigm, $\bm{\Theta}$ ensures the convergence points of different online models are close by each other.

Assuming to have a distribution of specific tasks, the derivation of meta-model $\bm{\Theta}$ follows this general procedure:
\begin{enumerate}
	\item Initialize $\bm{\Theta}$; 
	\item Randomly sample a task from the task distribution;
	\item Perform $K$ steps of gradient descent updates on the task, starting from $\bm{\Theta}$, so to obtain a new representation, denoted $\bm{W}$;
	\item Update $\bm{\Theta}$, that is,  $\bm{\Theta} \leftarrow \bm{\Theta} + \lambda(\bm{W}-\bm{\Theta})$, where $\lambda$ is the learning rate ($0<\lambda\leq1$);
	\item Repeat steps $2-4$ until $\bm{\Theta}$ is found to be optimal by the adopted optimization routine. 
\end{enumerate}
Once derived with the above procedure, meta-model $\bm{\Theta}$ can be used as a starting point for finding any specific model that suits a newly encountered task, by only using a small amount of experience collected on this new task~\cite{finn2017model, nichol2018first}. 

In our scenario, the learning tasks of FALCON are the different network conditions it may encounter and it should adapt to by deriving specific scheduling policies. In particular, we consider packet loss rate, mean RTT, and RTT variation rate of the available paths as indicators of network conditions. Moreover, FALCON adopts a DQN in the online learning module to derive its scheduling policies. Therefore, $\bm{\Theta}$ is defined as the initial set of parameters of the deep neural network used by DQN in the online learning module. Among common gradient descent approaches, we apply the mini-batch gradient descent rather than the stochastic gradient descent to cater for the use of DQN.

Note that the creation of a unique meta-model for all possible network conditions may require a significant increase of the number of gradient steps ($K$) needed to converge to an optimal $\bm{\Theta}$. Therefore, FALCON does not create a unique meta-model for representing all network conditions, but instead creates a set of meta-models, i.e.,  $\bm{\Theta}_{\mathsf{S}}$ (subscript $\mathsf{S}$ stands for set). Each meta-model in $\bm{\Theta}_{\mathsf{S}}$ is then created so to cover only a partial range of possible network conditions. For example, assuming to have two available paths, the $x$-th meta-model in $\bm{\Theta}_{\mathsf{S}}$, i.e., $\bm{\Theta}_{x}$, covers the range where, on path $1$, packet loss rate is between $[a, b]$ \%, mean RTT is between $[c, d]$ ms, and RTT variation rate is between $[e, f]$ \%, and similar bounds are defined for path $2$.            

The number of meta-models and the ranges of network conditions covered by different meta-models are predefined, as reported in Section \ref{setup}. In the following, $\bm{R}_{\mathsf{S}}$ denotes the set of ranges on which the meta-models operate.     
As shown in Algorithm~\ref{algorithm_O2}, the offline learning module updates $\bm{\Theta}_{\mathsf{S}}$ with a predefined 
update interval, i.e., $T_\mathsf{upd}$. To do so, it first collects experience on the currently deployed policy (state, action, reward, as defined in Section \ref{elements}) and on current network conditions (packet loss rate, mean RTT, and RTT variation rate), denoted $\bm{Exp}$, from the online learning module. 
Then, using the set of network conditions in $\bm{Exp}$, and comparing them with $\bm{R}_{\mathsf{S}}$, the offline learning module updates the corresponding meta-models in $\bm{\Theta}_{\mathsf{S}}$, following the procedure described above.

\vspace{1mm}
\noindent \textbf{\textit{Online learning module:}} This module runs continuously for deriving the scheduling policies to use under different network conditions. FALCON uses a change point detection mechanism to trigger the selection of the meta-model that covers the new conditions, and the derivation of a new policy leveraging the selected meta-model.  

Change point detection is thus an important aspect in FALCON, 
and it is particularly important in the wireless scenarios it faces, since these scenarios often result in high dynamicity and network changes, e.g., handovers. Intuitively, one could fix a detection interval and monitor the statistics of  network conditions in such an interval. Then, if the difference of statistics exceeds a threshold, a change in network condition is detected. How to setup the detection interval is, however, not trivial: if the interval is too short, the change detection may be affected by short-term noise; if it is too long, actual changes might be lost. A similar problem exists for setting up the threshold that identifies an actual change in network conditions. In short, a hard-coded setup of detection interval and threshold is not a viable approach.  

Therefore, since both gradual and sudden network condition changes are expected to happen in dynamic and heterogeneous networks, e.g., 5G mmWave~\cite{narayanan2020first}, we leverage the drift theory~\cite{gama2014survey} to observe the variability of network conditions. In particular, FALCON adopts the well-known Bayesian change point detection algorithm~\cite{adams2007bayesian} for monitoring loss rate and RTT on the available paths. On the one hand, the RTT is a continuous signal, and thus it can be used as is in the Bayesian change point detection algorithm; on the other hand, packet loss is a binary information (i.e., a packet can be either lost or not lost). To tackle this aspect, FALCON counts the packet losses over groups of packets, thus moving from a Bernoulli distribution to a binomial distribution of packet losses, and finally obtaining a relatively continuous signal. 

As shown in Algorithm~\ref{algorithm_O2}, upon detection of a change in network conditions, the online learning module selects the meta-model in $\bm{\Theta}_{\mathsf{S}}$ that covers the range that the current network condition belongs to by checking the network conditions against $\bm{R}_{\mathsf{S}}$ over a short period of data transmission. When localizing the current network condition, there might exist bias due to the noise. Recall that the meta-model covers a range of link characteristics in FALCON, which reasonably tolerates these biases. 
Once the meta-model is selected, FALCON performs a $K$-step fine-tuning of the meta-model and, via DQN, derives the scheduling policy to adopt. Finally, FALCON deploys and uses the new policy until a new change is detected. Since the learning agent does not grow its knowledge base from null rather from the shared knowledge, the cost of adaptation is fairly small as shown in Section~\ref{emulation}.

\subsection{Learning Elements}
\label{elements}
As anticipated in Section \ref{overall}, FALCON uses a DQN architecture for deriving the policy at run-time, and exploits the meta-learning paradigm to speed up such derivation while preserving accuracy. Therefore, the entire framework is a MDP that FALCON solves via \emph{meta-learning plus DQN}. In the following, we provide a few more details on the learning elements of the overall framework.

\vspace{1mm}
\noindent\textbf{\textit{State space:}} The state in a MDP is the information observed by a learning agent on the status of the environment that the agent is facing during the learning process. In our scenario, FALCON is the learning agent and the environment state is defined via transport layer parameters of the available paths, i.e., CWND, number of Inflight Packets (InP), Send Window (SWND) and RTT. The first three features are normalized by RTT to impose a tight connection to the throughput, that is the reward FALCON obtains while operating, as introduced later.

\vspace{1mm}
\noindent\textbf{\textit{Action space:}} This set includes the actions FALCON can select when it deploys a scheduling policy, and upon which it gets a reward. In our scenario, the available actions depend on the number of available paths. In this work, we mostly consider two available paths, which is a common assumption in 5G multi-connectivity scenarios~\cite{TS23501}. However, the action set of FALCON can be naturally extended and include more paths. Hence, by taking an action, FALCON decides about the path to use for exchanging a data packet. 
In the context of multipath scheduling, this indicates that the action set can be different when a path is congested or not congested,  which complicates the learning agent. We choose to naturally inherit this information from the state while appointing the path in a straightforward manner regardless of the path's current congested status.    

\vspace{1mm}
\noindent\textbf{\textit{Reward function:}} As common in a MDP, FALCON aims at maximizing a so-called discounted return, where the instantaneous rewards, corresponding to the throughput obtained upon selection of a path, are cumulated after being discounted via a so-called discount factor, that can be interpreted as the interest of the scheduler in maximizing short vs. long-term return. 
The employment of the discounting factor ensures that the impact of the current action decreases over time. 

\vspace{1mm}
\noindent\textbf{\textit{RL algorithm:}} As anticipated above, FALCON uses DQN in the online learning module for deriving the scheduling policy. DQN is a well-known model-free algorithm that does not require any knowledge of the state transition probability distribution and the reward function. On the contrary, it just needs to observe the instantaneous rewards obtained when actions are selected, and the corresponding transition across states. While considering complexity as a primary factor, we select DQN also due to its popularity, which makes possible a direct comparison with other DQN-based state-of-the-art schedulers, as introduced in Section \ref{schedulers}. However, it is worth to mention that FALCON is based on a rather flexible framework and can be thus easily extended toward the adoption of other algorithms in the online learning module.  

\vspace{1mm}
\noindent\textbf{\textit{Exploration vs. exploitation:}} Due to the presence of the information exchange between the offline and online learning modules, FALCON requires a certain degree of balance between exploration and exploitation. Hence, it adopts a fixed  $\epsilon$-greedy exploration mechanism, with $\epsilon$ not decaying over time. In particular, a relatively large value of $\epsilon$, denoted $\epsilon_l$, is used when setting up these initial meta-models, aiming for higher sampling efficiency; on the contrary, a relatively small value of $\epsilon$, that is, $\epsilon_s$, is used when the meta-models are continually updated and also when the selected meta-model is fine-tuned to derive the scheduling policy. 

\vspace{1mm}
\noindent\textbf{\textit{Synchronous vs. asynchronous learning:}} In the original proposal of DQN and common deep RL paradigms, the interaction with the environment and the update of the neural network happen in a synchronous fashion. However, these synchronous operations do not work well in a real system where there usually exist either soft or hard real-time requirements. For example, in our case, the online update of the neural network could block the scheduling routine in the communication stack. We thus employ asynchronous updates~\cite{gu2017deep}, implemented by using a separate process for the online learning: A network process is in charge of the data collection and performs the scheduling while a trainer process is in charge of neural network updates based on the collected data.

\section{Experimental setup}
\label{setup}
In this section we present the experimental setup, including the configuration of FALCON, the selected baseline multipath scheduling algorithms, and the experimental environment.  

\subsection{Configuration of FALCON}
We implement the learning components of FALCON based on \texttt{keras-rl}~\cite{plappert2016kerasrl}, a popular deep reinforcement learning library. We employ a fully connected neural network with three hidden layers and a rectified linear activation function (ReLU) as the activation function. The learning rate of the neural network is 0.001, while $\epsilon_l$ and $\epsilon_s$ are 0.3 and 0.1, respectively. The mini-batch size is 32 and $K$ is 16. For the ranges of network conditions that the meta-models cover over each path, we implement that the packet loss rate can be in between $[0, 1)$\%, $[1, 5)$\%, and $[5, 100]$\%; mean RTT can be in between $[0, 50)$ ms, $[50, 200)$ ms, and $[200, +\infty)$ ms; the ratio of RTT deviation to mean RTT can be in between $[0, 40)$\%, $[40, 80)$\%, and $[80, +\infty)$\%. Thus, one path, by combination, can have 27 different coarse-grained states and two paths, by combination, can have 729 different coarse-grained states. Accordingly, the number of meta-models in total is 729. The online experiences are periodically written to a comma-separated values (CSV) file and the neural networks representing the meta-models are saved into the Hierarchical Data Format version 5 (HDF5) file.  

We set the specific parameters of FALCON based on our experimental analyses in Section \ref{indepth::study}. We believe these are reasonable design choices in practice, and note that our analysis enables tuning this parameter to accommodate other scenarios. Unless stated otherwise, our evaluation of FALCON uses these default values. 
 
\subsection{Configurations of the  protocol stack}
At the transport layer, we perform our analysis using MPQUIC due to the increasing interest in QUIC-based applications. Accordingly, the QUIC is originally implemented within \texttt{quic-go}~\cite{Lucas2019} and, based on which, one of the earliest versions of MPQUIC is implemented and adopted in this work. Further we use the default multipath congestion control algorithm in the adopted MPQUIC code base, i.e., Opportunistic Linked-Increases Algorithm (OLIA)~\cite{khalili2013mptcp}.

At the application layer, 
we perform both bulk transfer and web download to evaluate the aggregation capability of the multipath schedulers. For the bulk transfer, each experiment run performs an \texttt{HTTP GET} request for a file of 2 MB, and records the download times. For the web download, we consider web pages from different websites including Google, Github, and Stackoverflow, as shown in Table~\ref{tab_parameter}, and we record the download times. Transport layer state variables are reset before each request. To ensure statistically significant results, for each path configuration, we repeat the experiments 120 times for each multipath scheduler.

\begin{table}[t]
\caption{\footnotesize{ \uppercase{Emulation Parameters for the static scenario, collected from measurements in the literature.}}}
\centering
\begin{tabular}{cccc}
\Xhline{2\arrayrulewidth}
\textbf{Parameter} & \multicolumn{3}{c}{\textbf{Link Technology}} \\ 
\hline
 & \begin{tabular}[c]{@{}c@{}}\emph{5G}\\ \cite{narayanan2020first}\end{tabular} & \begin{tabular}[c]{@{}c@{}}\emph{4G}\\ \cite{narayanan2020first}\end{tabular} & \begin{tabular}[c]{@{}c@{}}\emph{WLAN}\\ \cite{sheshadri2017packet}\cite{pei2016wifi}\end{tabular} \\ \cline{2-4}
Bandwidth {[}Mbps{]} & $\sim$ 1100 & 140 & 30 \\
Round trip time (RTT) {[}ms{]} & 27.4 $\pm$ 6.4 & 29.2 $\pm$ 4.8 & 20.0 $\pm$ 10.0 \\
Packet Loss Rate [\%] & 0.1 & 0.1 & 0.7 \\ 
\Xhline{2\arrayrulewidth}
\end{tabular}
\label{tab_parameter}
\end{table}

\begin{table}[t]
\caption{\footnotesize{ \uppercase{\rev{Parameters of selected websites for the web download test.}}}}
\centering
\begin{tabular}{ccc}
\Xhline{2\arrayrulewidth}
\textbf{Website} & \multicolumn{2}{c}{\textbf{Parameters}} \\
\hline
 & \begin{tabular}[c]{@{}c@{}}\emph{Number of Objects}\end{tabular} & \begin{tabular}[c]{@{}c@{}}\emph{Size} \end{tabular}  \\ \cline{2-3}
Google & 8 & 1.3 MB  \\
Github & 61 & 4.5 MB  \\
Stackoverflow & 120 & 7 MB \\ 
\Xhline{2\arrayrulewidth}
\end{tabular}
\label{tab_parameter}
\end{table}

\subsection{Benchmark Algorithms} 
We select \textbf{minRTT} and \textbf{BLEST} as the representative algorithms of the schedulers based on predefined rules, for two reasons: 1) the strategies they exploit consider the challenges originating not only from the homogeneous networks but also the heterogeneous networks; 2) recent evaluations show that they either perform similarly  or better than other schedulers belonging to the same category (e.g., RR and ECF)~\cite{WuCommag21,wu2020peekaboo}.

\begin{figure*}[t]
	\centering 
	\subfigure[\footnotesize Static: 4G and 5G] {\includegraphics[draft=false,width=0.49\columnwidth]{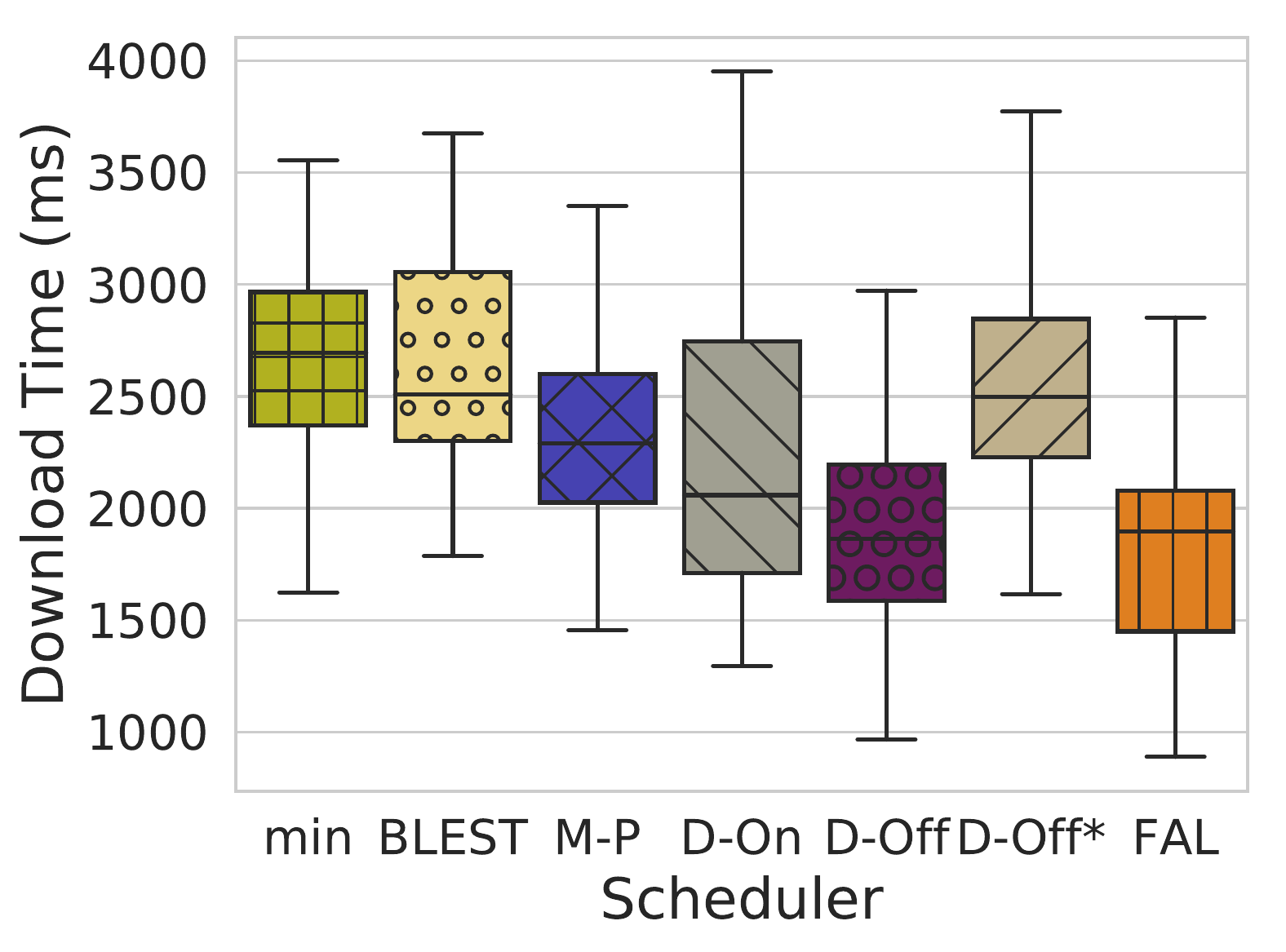}} 
	\subfigure[\footnotesize Static: 4G and WLAN] {\includegraphics[draft=false,width=0.49\columnwidth]{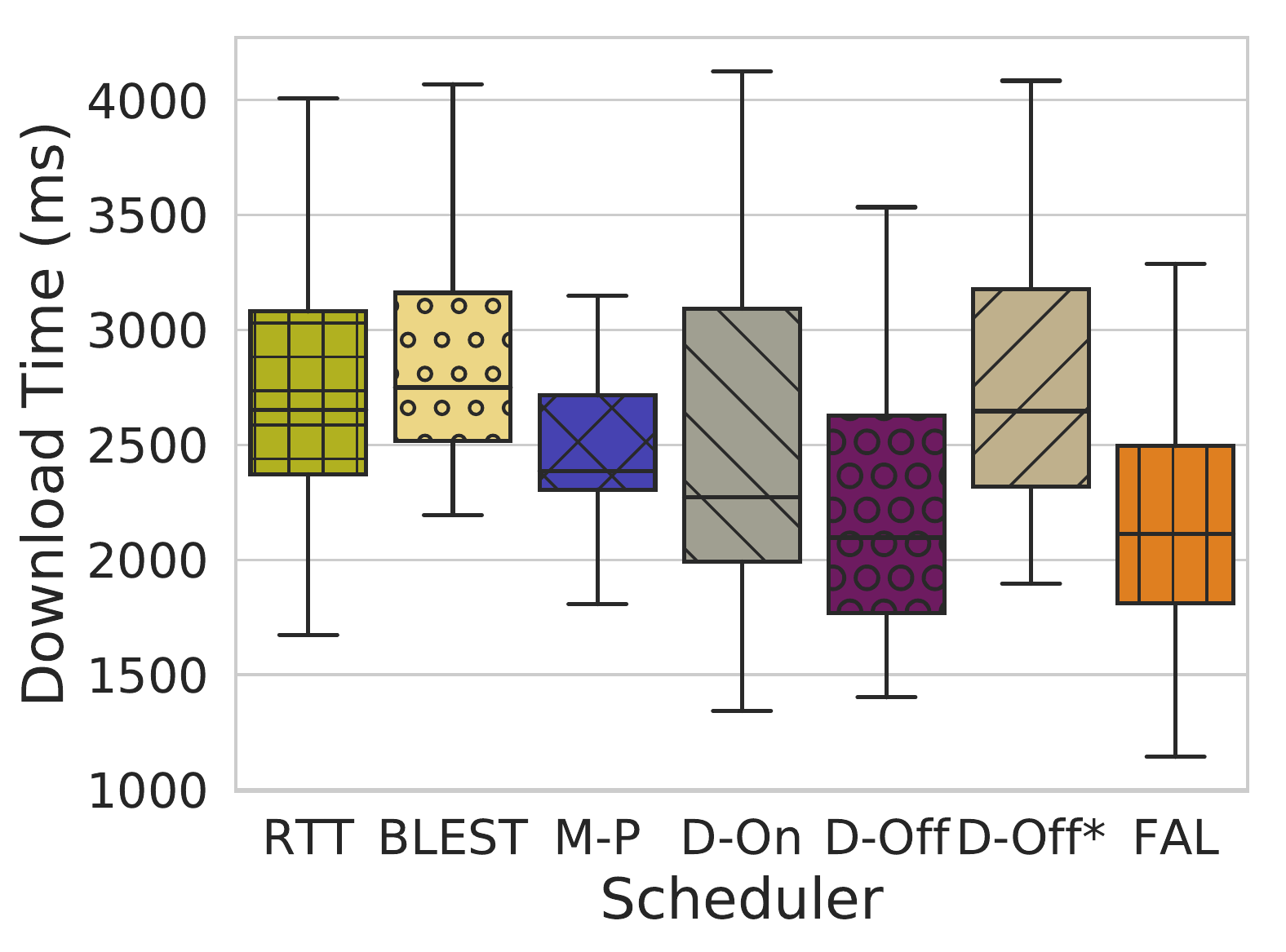}} 
	\subfigure[\footnotesize Static: 5G and WLAN] {\includegraphics[draft=false,width=0.49\columnwidth]{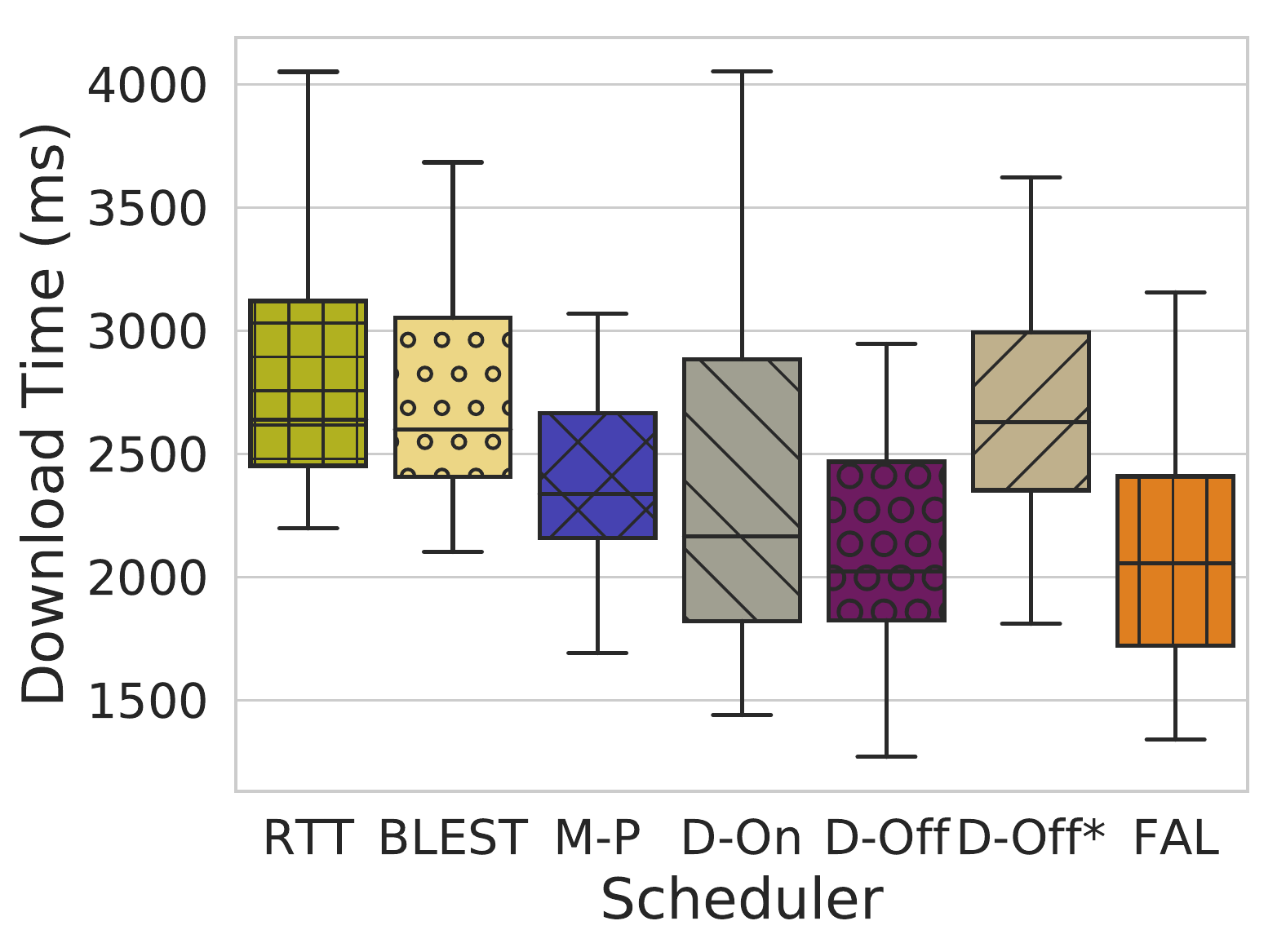}}
	\subfigure[\footnotesize Driving: 5G and 5G] {\includegraphics[draft=false,width=0.49\columnwidth]{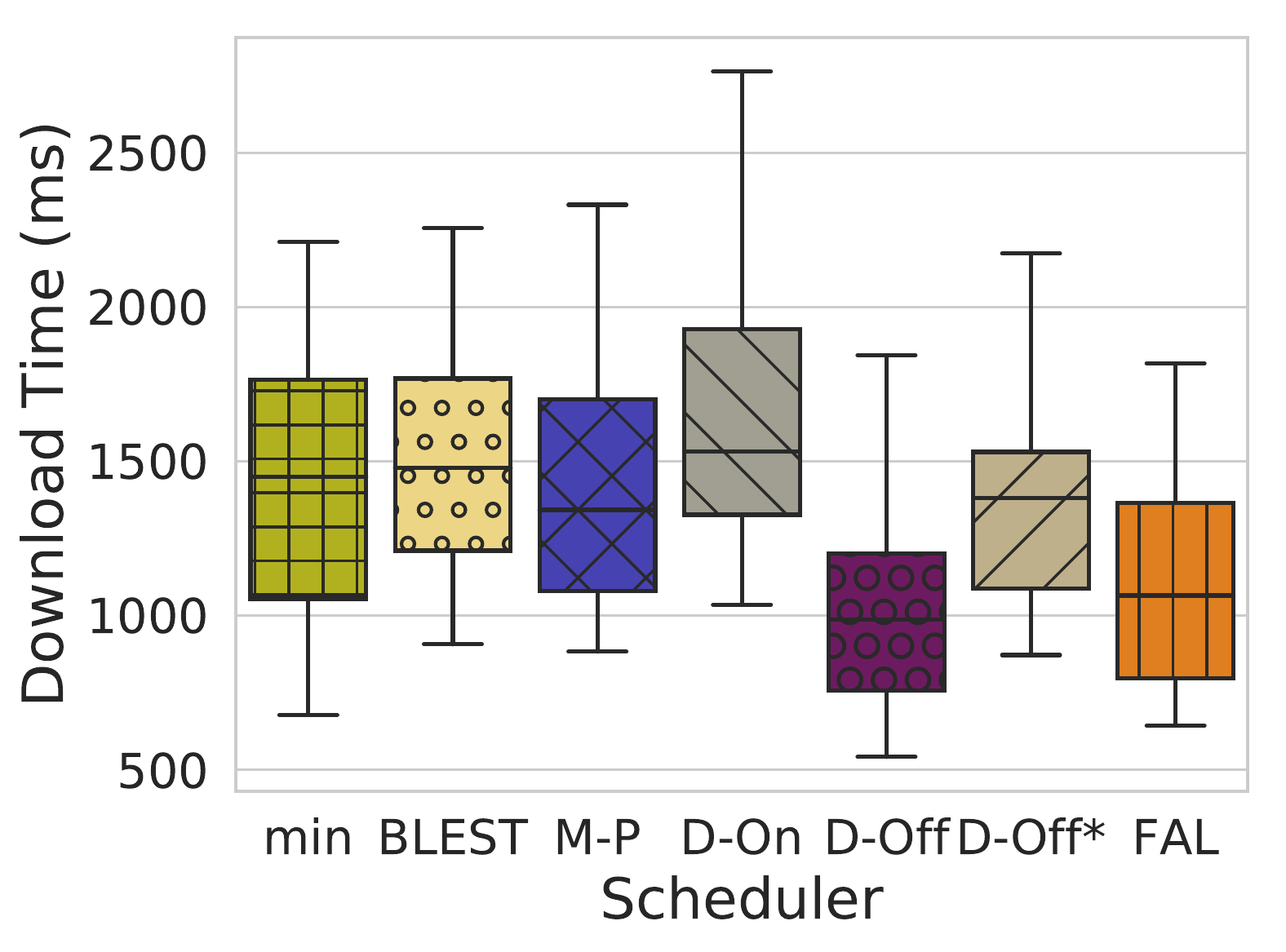}} 
	\caption{Performance of FALCON and other multipath schedulers in the static and mobile network conditions over 4G, 5G, and WLAN.}
	\label{fig_dqn full on off}
\end{figure*}

For the offline learning-based schedulers, as there is not any multipath scheduling algorithms in the literature, we refer to a recent implementation in the area of ABR streaming~\cite{akhtar2018oboe}, and implement a DQN-based multipath scheduling algorithm, named \textbf{DQN-Off}. In the implementation, while keeping the concept of offline training, we utilize the off-policy characteristics of DQN rather than a simulated environment to achieve this goal. That means, DQN off-policy is able to learn from information retrieved from past experience rather than direct interaction with the environment. Moreover, we use the same learning elements used by FALCON, described in Section \ref{elements}. 

For Type-I online learning based schedulers, we refer to the \textbf{M-Peekaboo} algorithm that can exploit the linUCB and stochastic adjustment algorithm to learn the scheduling policy \cite{WuCommag21}. For Type-II online learning based schedulers, the DQN-based scheduler designed in~\cite{rosello2019multi} fails to provide clear performance gains over schedulers based on predefined rules, while~\cite{zhang2019reles} shows performance gains. However, the authors of~\cite{zhang2019reles} do not disclose the source code, and we do not have enough information to reproduce the work. Hence, we use all the information that can be extracted from~\cite{rosello2019multi} and~\cite{zhang2019reles}, and implement a DQN-based online multipath scheduler, i.e., \textbf{DQN-On}. In particular, within the framework provided by~\cite{rosello2019multi}, we exploit the same state space, action space, and reward function used by FALCON, to have a higher granularity representation in the constructed MDP. 
Then, we also adopt the asynchronous online update mechanism used in~\cite{zhang2019reles}, originally proposed in~\cite{gu2017deep}, to speed up the training time of DQN when applied to real-world applications. Although \cite{zhang2019reles} refers to this asynchronous online update mechanism as a combination of online and offline learning, we highlight that this is in essence an implementation choice of inter-process communication, and thus it differs from the concept of online and offline learning defined in this work. 

\subsection{Experimental environment}
\label{setup:environment}
We perform experiments over both emulated and real-world \textit{urban canyon} environments. In both cases, we consider two scenarios: \emph{static}, where we assume the user is stationary, and \emph{mobile}, where the user is walking and/or driving a vehicle. 

In the emulated experiments, aiming at a controlled but realistic evaluation, we leverage link characteristics derived from real measurements, with both network traces and statistical values, as shown in Table~\ref{tab_parameter}. The environment is emulated using \texttt{Mininet}~\cite{handigol2012reproducible}. Regarding path characteristics (i.e., bandwidth, latency, and packet loss), we use values measured against a content-server close to the radio infrastructure, mimicking 5G edge deployments.  
In the static scenario, we showcase multipath transport between 4G and 5G paths and between 4G and WLAN paths as well as between 5G and WLAN paths. Our motivation to evaluate all these options is due to the proposed ATSSS architecture from 3GPP. In the mobile scenario, we showcase multipath transport over two 5G networks in a driving scenario~\cite{narayanan2020first}.

\begin{figure*}[t]
	\centering 
	\subfigure[\footnotesize 4G and 5G] {\includegraphics[draft=false,width=0.65\columnwidth]{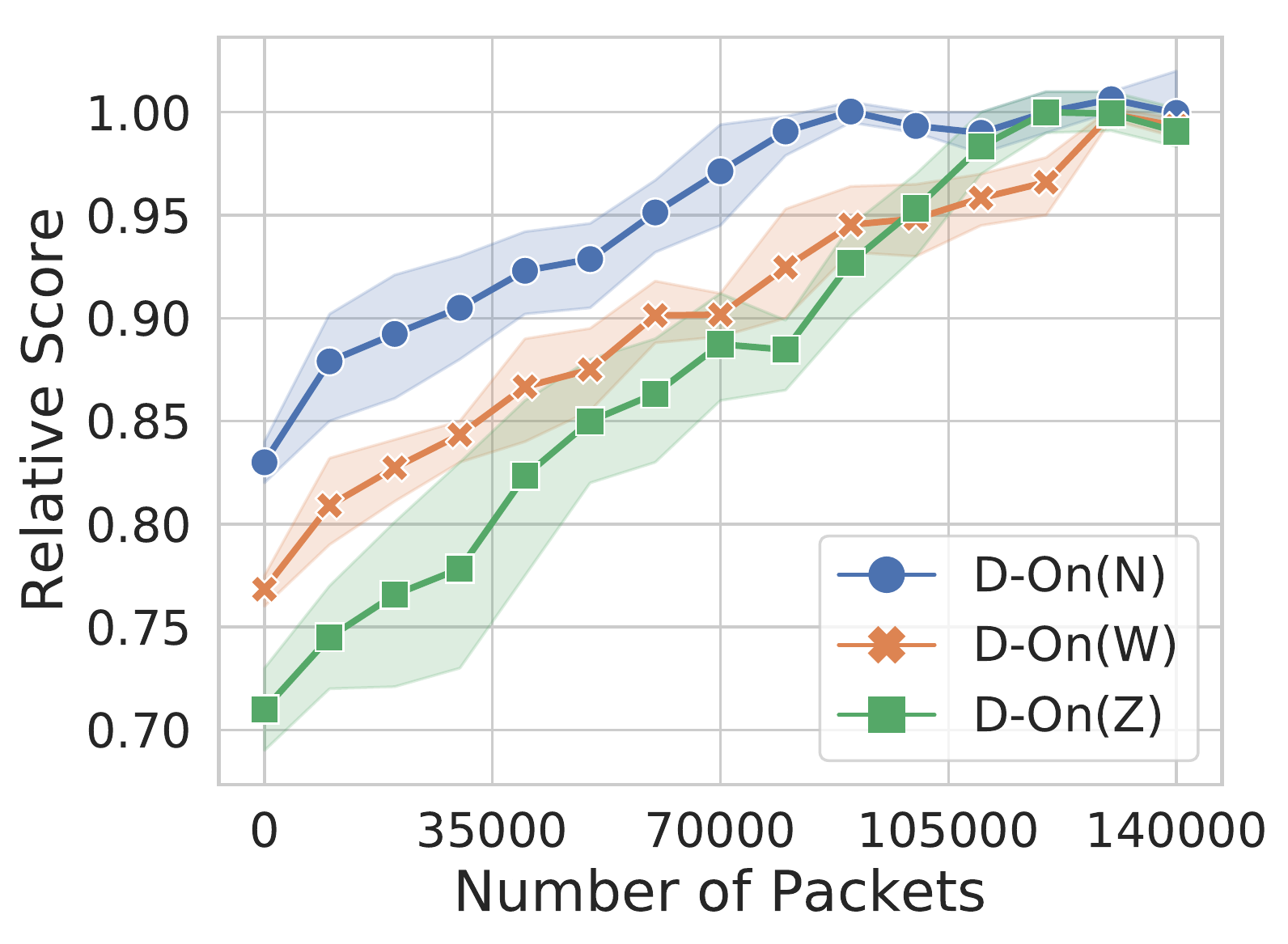}} 
	\subfigure[\footnotesize 4G and WLAN] {\includegraphics[draft=false,width=0.65\columnwidth]{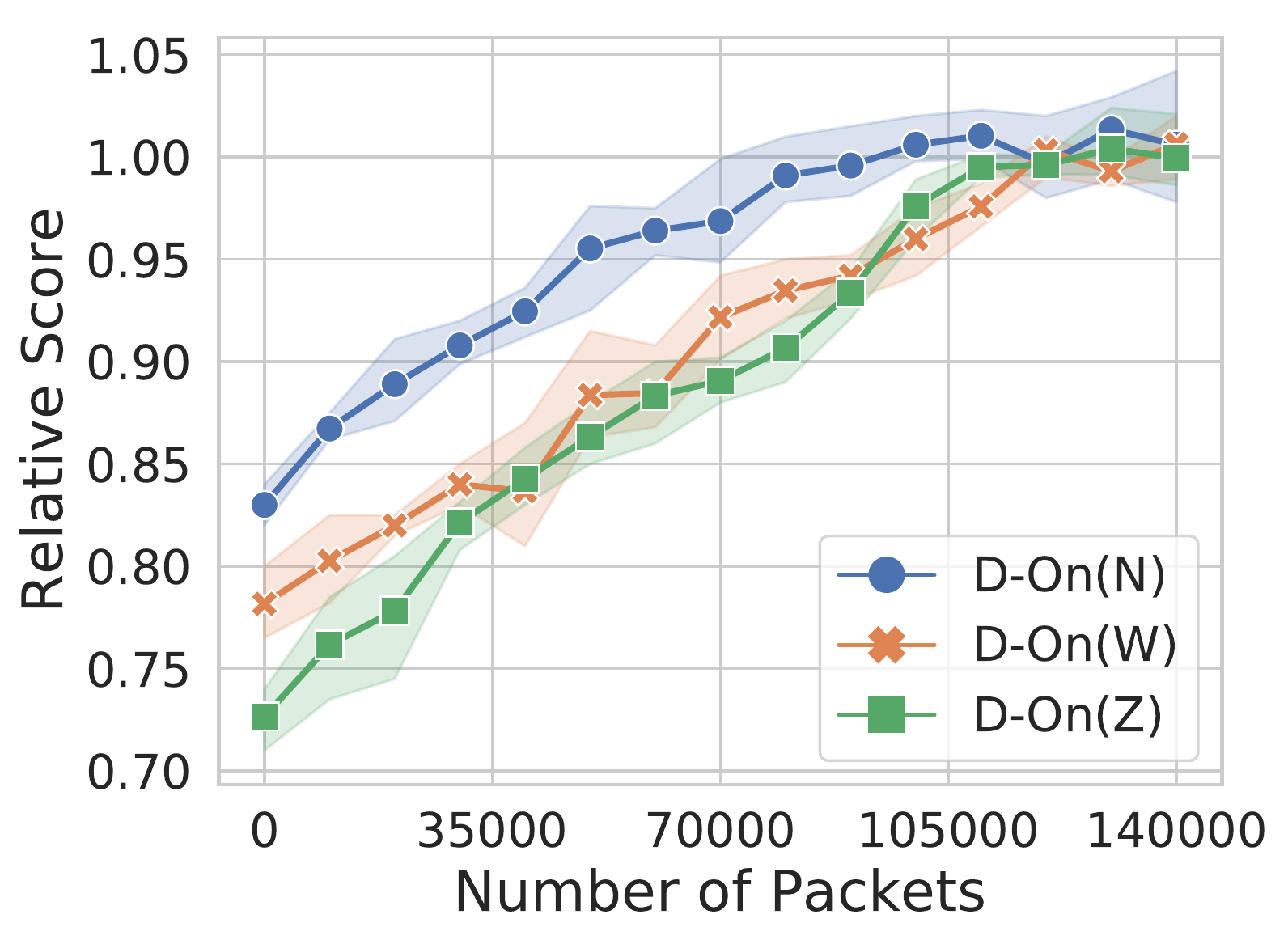}} 
	\subfigure[\footnotesize 5G and WLAN] {\includegraphics[draft=false,width=0.65\columnwidth]{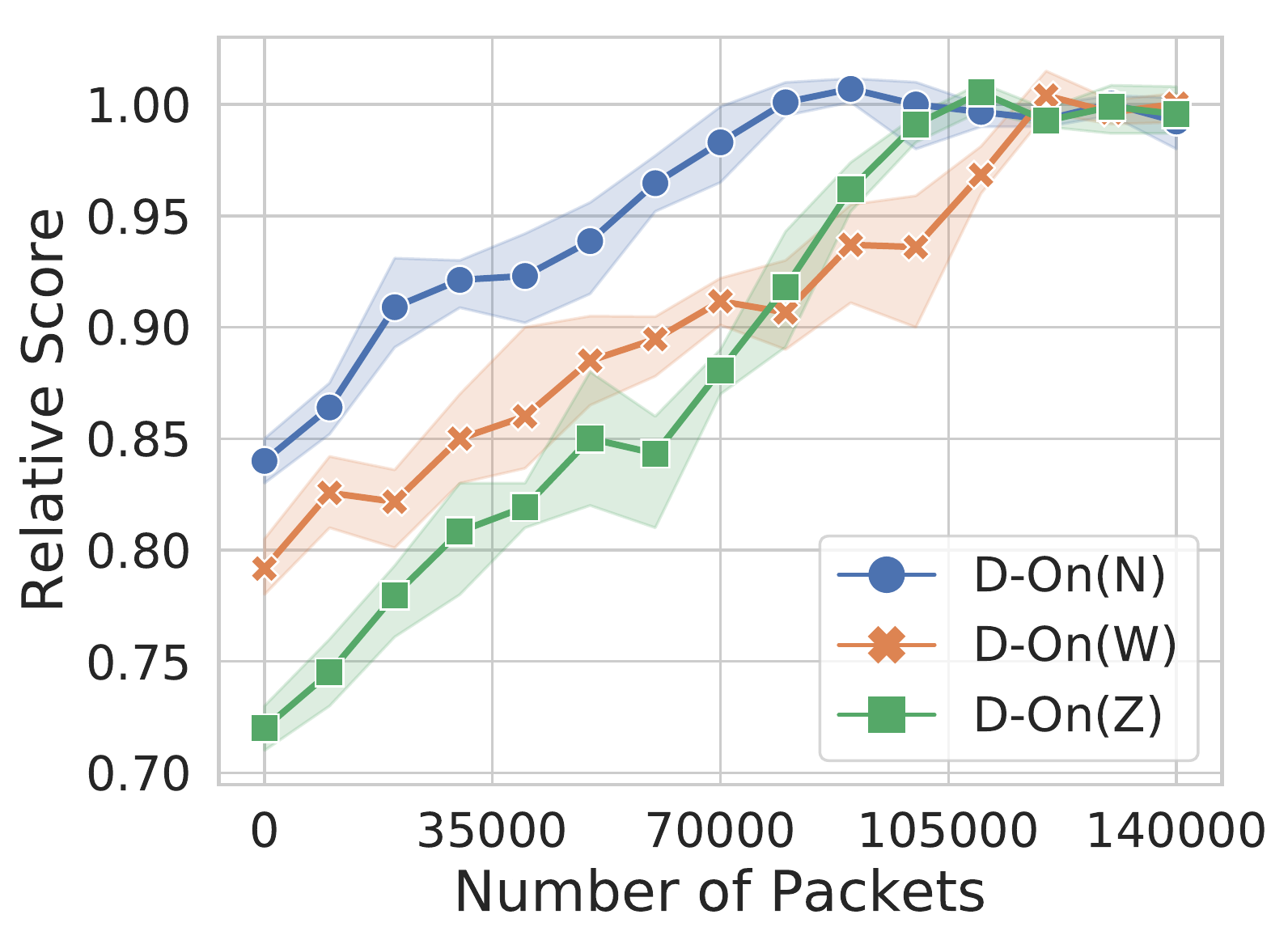}}
	\caption{Convergence test of DQN-On in different configurations over 4G, 5G, and WLAN.}
	\label{fig_dqn cprogress}
\end{figure*}

In the real-world experiments, we analyze multipath transport between 5G and WLAN in the static scenario. In the mobile scenario, we showcase multipath transport over 4G and 5G in a driving test.

\section{Emulated Experiments} 
\label{emulation}
In this section, we compare the performance of FALCON with state-of-the-art multipath schedulers in a wide range of emulated experiments.
We use the bulk transfer case to analyze the performance of the multipath schedulers in static and mobile scenarios  (Section~\ref{emulate::static}), and provide more insights on the behavior of FALCON and other schedulers in terms of how quickly they adapt to time-varying network conditions (Section~\ref{emulate::closerlook}). Finally, we further validate the robustness of FALCON in web download scenarios    (Section~\ref{emulate::extend}).  

\subsection{Performance in Static and Mobile Scenarios} 
\label{emulate::static}

We first evaluate the performance of different multipath schedulers in static and mobile scenarios. We focus on the analysis of schedulers based on learning while keeping schedulers based on predefined rules as a reference. For the schedulers based on learning with online approach (FALCON, DQN-On, M-Peekaboo), we assume: (i) they were not trained over the examined network conditions beforehand, and (ii) they have no buffered online data at the beginning of each experiment. On the other hand, to directly compare the impact of an approach with offline pre-knowledge, we assume that DQN-Off was trained over the examined network conditions beforehand. 

Figure~\ref{fig_dqn full on off} presents the performance of different schedulers under different scenarios as described in Section~\ref{setup:environment}. For the static case (Figure~\ref{fig_dqn full on off} a-c), we observe that all the schedulers based on learning (FALCON, DQN-On, DQN-Off, M-Peekaboo) outperform the schedulers based on predefined rules (minRTT, BLEST) with up to 34.5\% shorter median download time. Concerning schedulers based on learning, we observe that schedulers utilizing deep learning, including FALCON, DQN-Off, and DQN-On, outperform M-Peekaboo with up to 19.3\% shorter median download time. As all the schedulers adapt to the presented static network condition, this indicates that applying a model of high complexity is beneficial for improving the adaptation accuracy. We also observe that the performance of FALCON is similar to DQN-Off and clearly better than DQN-On. This indicates that FALCON can adapt faster compared to DQN-On thanks to its few-shot online learning, which allows to achieve the same accuracy as DQN-Off.  

Note that we assumed DQN-Off was trained over the examined network conditions beforehand and is able to deploy an accurate model without the additional cost of online learning. However, it is rarely the case that the online data is fully and well-aligned with the offline data under realistic settings. To capture this effect, we consider the scenario, where the model obtained during training deviates from the current network conditions by only a 5\% decrease in terms of the RTT variation and loss rate of the paths. We denote DQN-Off under these new deviated network conditions as DQN-Off$^*$. We observe that there is a significant performance drop of DQN-Off$^*$ compared to DQN-Off with up to 34.5\% longer median download time. Its performance is similar to the schedulers with pre-defined rules. This indicates that DQN-Off lacks the ability to adapt, hence negatively impacting its practicality under realistic settings.

Next, we evaluate the performance of FALCON and baseline schedulers in a mobile scenario. We illustrate the performance of different schedulers in the trace-driven mobile network conditions in Figure~\ref{fig_dqn full on off}(d). We observe that the performance gain of M-Peekaboo over the schedulers based on predefined rules decreases compared to that of the static case, since it does not adapt fast enough to the less predictable changes of network conditions. M-Peekaboo, however, outperforms DQN-On by an 18.9\% shorter median download time, since it has a more lightweight learning mechanism and, thus, a shorter adaptation time. The adaptation time of DQN-On is quite long because of the intrinsic slow convergence time of DQN, which we investigate separately in Section~\ref{emulate::dqnconv}. Thanks to the few-shot online learning, FALCON still clearly outperforms the online learning based schedulers, reaching a 16\% shorter median download time compared to M-Peekaboo. DQN-Off performs slightly better than FALCON as it does not have the cost of few-shot learning during frequent network condition changes. However, when we introduce a model deviation, as done for the static scenario, we observe again that DQN-Off$^*$ performance drops significantly, since it lacks the ability to adapt to the deviated network conditions.

\begin{figure*}[t]
	\centering 
	\subfigure[\footnotesize 4G and 5G] {\includegraphics[draft=false,width=0.65\columnwidth]{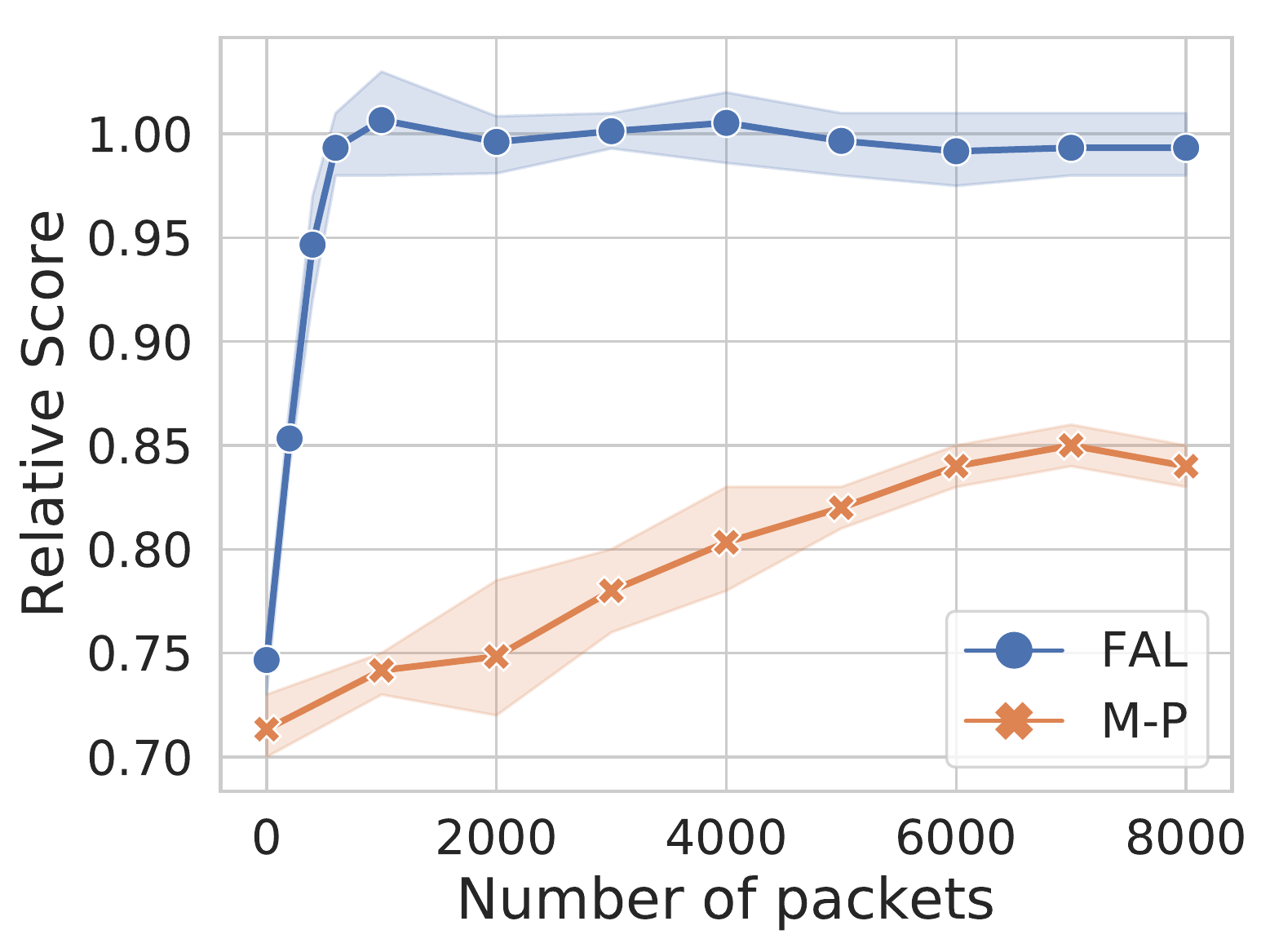}} 
    \subfigure[\footnotesize 4G and WLAN] {\includegraphics[draft=false,width=0.65\columnwidth]{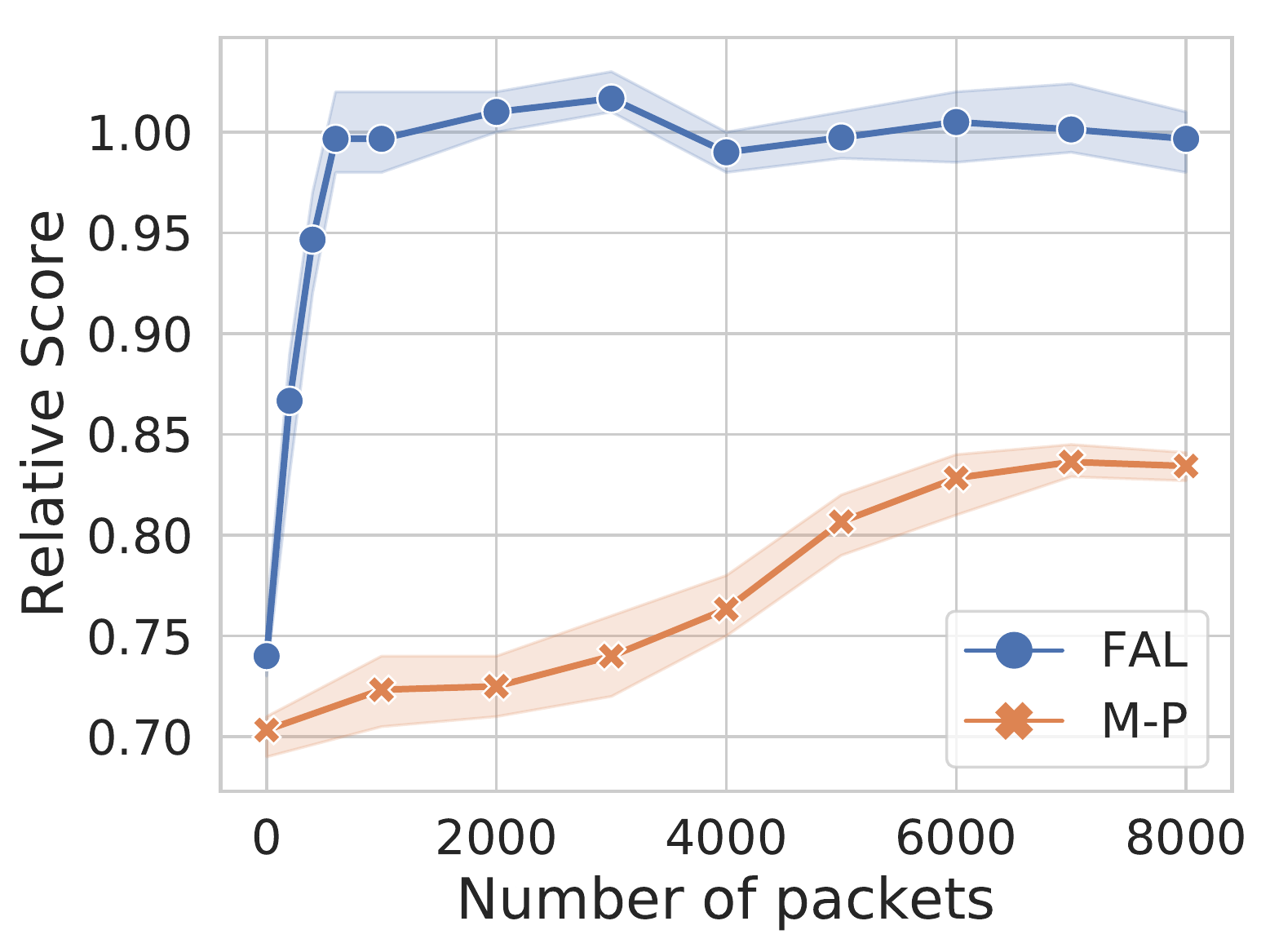}} 
    \subfigure[\footnotesize 5G and WLAN] {\includegraphics[draft=false,width=0.65\columnwidth]{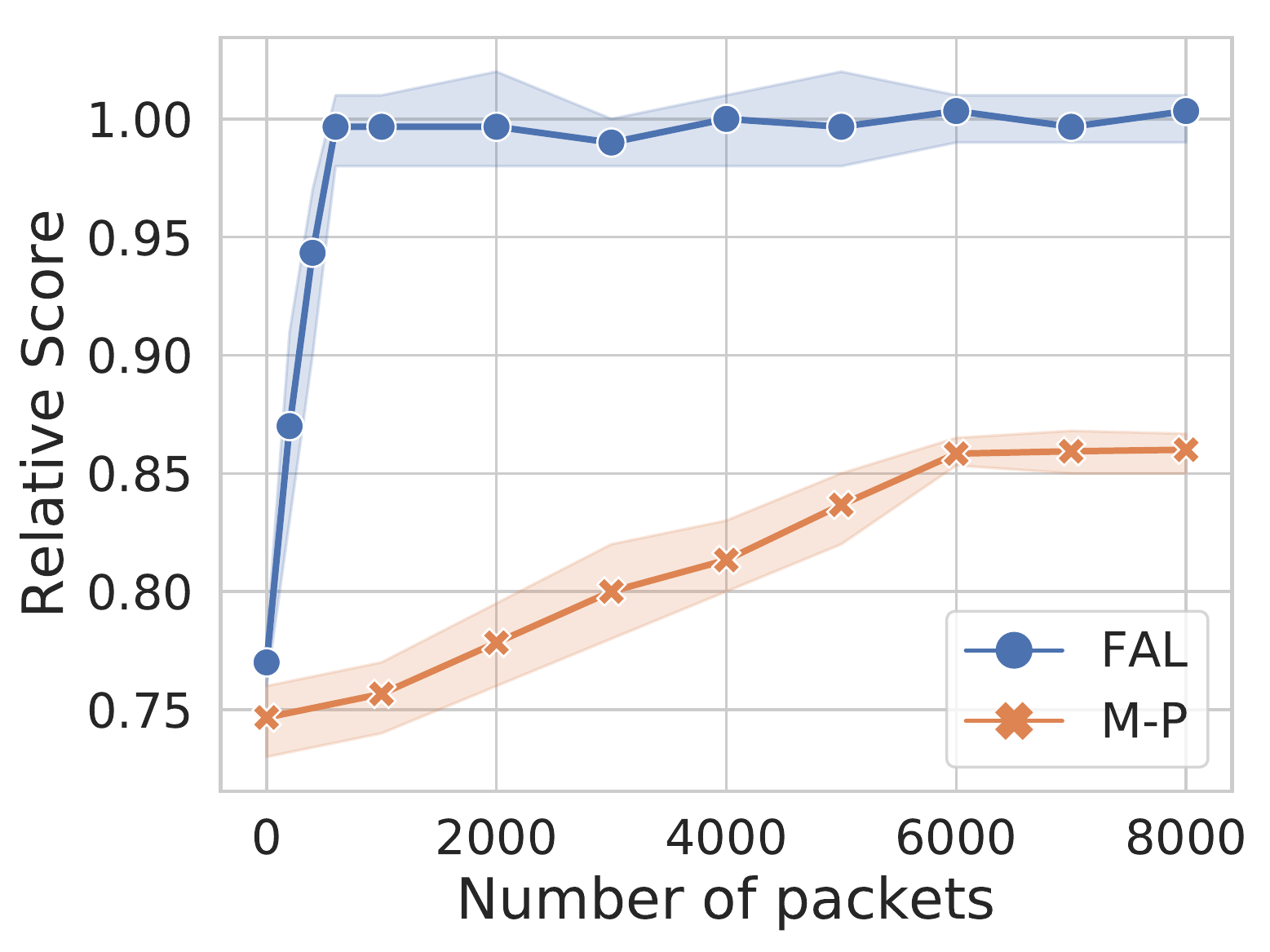}} 
	\caption{Convergence test of FALCON and M-Peekaboo over 4G, 5G, and WLAN.}
	\label{fig_dqn cprogressfalcon}
\end{figure*}

\subsection{A Closer Look into Adaptation Time}
\label{emulate::closerlook}
We now examine in depth the factors that impact the adaptation time for the schedulers based on learning with online adaptation, i.e, DQN-On, M-Peekaboo, and FALCON.

\subsubsection{Convergence Test}
\label{emulate::dqnconv}

We perform a convergence test to explore the convergence behavior of DQN-On, M-Peekaboo, and FALCON. We define the relative score as the ratio between the median file download time obtained by DQN-Off and the median file download time obtained by the scheduler under test. 
We use the relative score to illustrate how online learning algorithms evolve over time, thus, we evaluate this score as a function of the online learning cost, i.e., the number of online packets at the transport layer.
For each scheduler, we perform the test 10 times. For DQN-On, we also consider the impact of buffered online data due to previous training. Therefore, we not only investigate the convergence of DQN-On with zero buffered online data, i.e., no previous training, (denoted by DQN-On(Z)) but also with a narrow vs. wide range of buffered online data (denoted by DQN-On(N) and DQN-On(W), respectively). In particular, DQN-On(N) was trained beforehand over two network conditions that have, compared to current network conditions, a 3\% decrease or a 3\% increase of RTT variation and loss rate on the available paths. DQN-On(W) was instead trained over four network conditions having 3\% and 6\% decrease and increase of the same indicators, respectively. The experience on the network conditions is obtained by exploiting a learning budget of 100 packets exchanged under each of these conditions. 

Figure~\ref{fig_dqn cprogress} shows the results of the convergence test for DQN-On with different amount of buffered online data. The number of packets for reaching convergence is very different when we compare DQN-On and FALCON / M-Peekaboo; therefore we show the results of DQN-On in Figure~\ref{fig_dqn cprogress} and the results of FALCON / M-Peekaboo in Figure~\ref{fig_dqn cprogressfalcon}. 
We observe that DQN-On requires a large amount of data to converge, in the order of 100,000 packets. DQN-On(N) has a relatively higher score at the beginning and shows earlier convergence compared to DQN-On(W). This is due to two main reasons: first, the network conditions it has trained over have a higher similarity to the current network conditions; second, the total number of packets in its learning budget is smaller (200 for DQN-On(N) vs. 400 for DQN-On(W)), so in DQN-On(N) the online data from current network conditions dominates faster the buffered online data, ultimately speeding up adaptation. Similarly, DQN-On(W) has a relatively higher score than DQN-On(Z) at the beginning, due to the training beforehand, but DQN-On(Z) converges earlier than DQN-On(W), since it does not need to nullify the impact of buffered online data deviating from current network conditions. The analysis indicates that buffered online data that deviates from current conditions may harm convergence. Note that we allocated a small amount of online data that are relatively close to the current network conditions (deviations are within 6\%); In more realistic settings, even more data with wider deviated ranges could be buffered, leading to a continuous slow down of the adaptation time of DQN-On. 

Next, we illustrate the results of the convergence test for M-Peekaboo and FALCON in Figure~\ref{fig_dqn cprogressfalcon}.
We observe that FALCON and M-Peekaboo achieve convergence with an approximate learning cost of 600 and 6,000 packets, respectively. These values are much smaller than that of DQN-On. Meanwhile, the paradigms of FALCON and M-Peekaboo are free of the impact of the buffered online data. Further, we observe that M-Peekaboo has a relatively fast convergence speed but its relative score is lower than FALCON, due to its simpler learning model. 
On the other hand, by combining offline and online learning, FALCON not only converges faster but also achieves a higher score compared to M-Peekaboo.

\begin{figure}[t]
	\centering 
\includegraphics[draft=false,width=0.88\columnwidth]{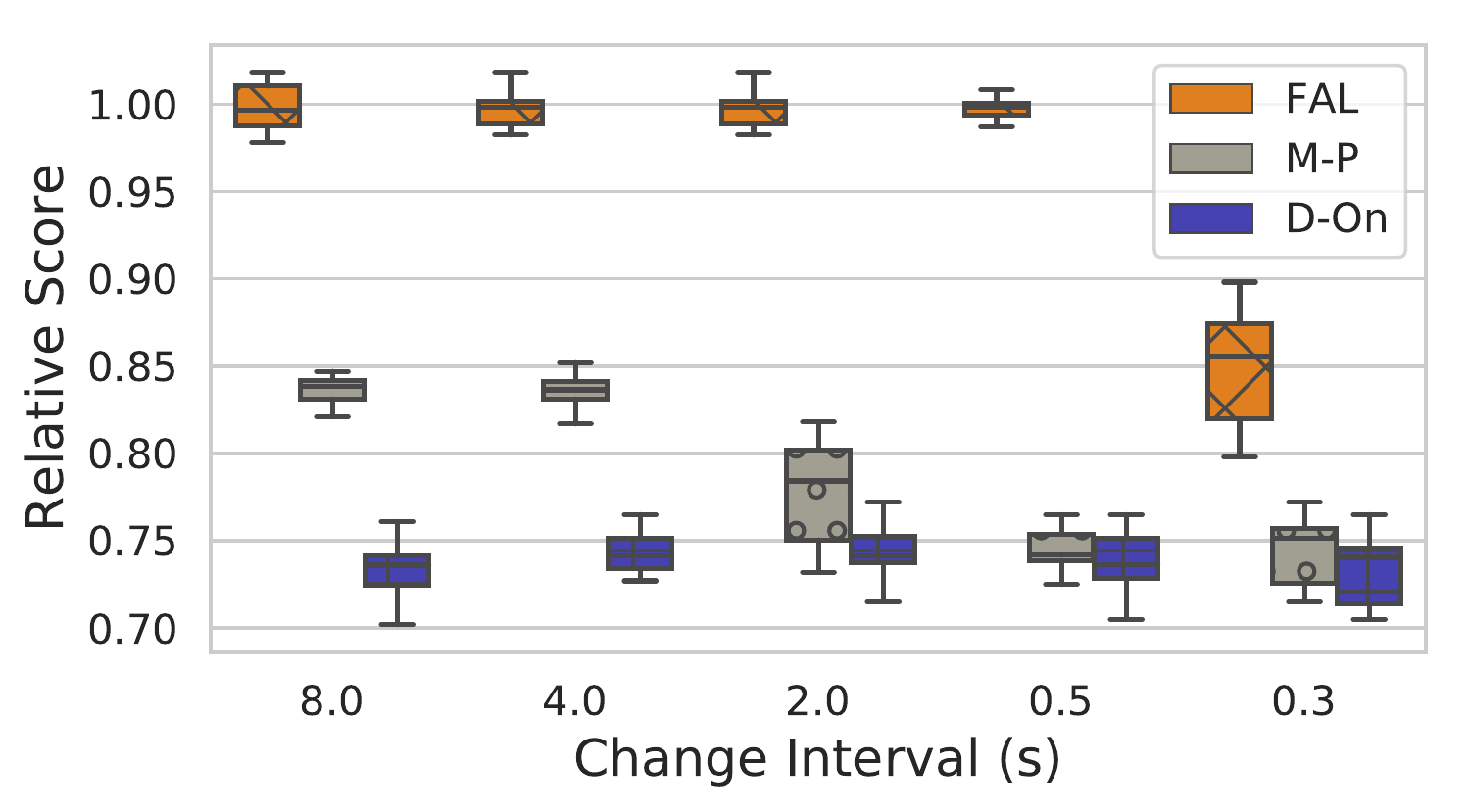}
	\caption{Stress test of FALCON, M-Peekaboo, and DQN-On.}
	\label{fig_dqn stress}
\end{figure}

\begin{figure}[t]
	\centering 
\includegraphics[draft=false,width=0.88\columnwidth]{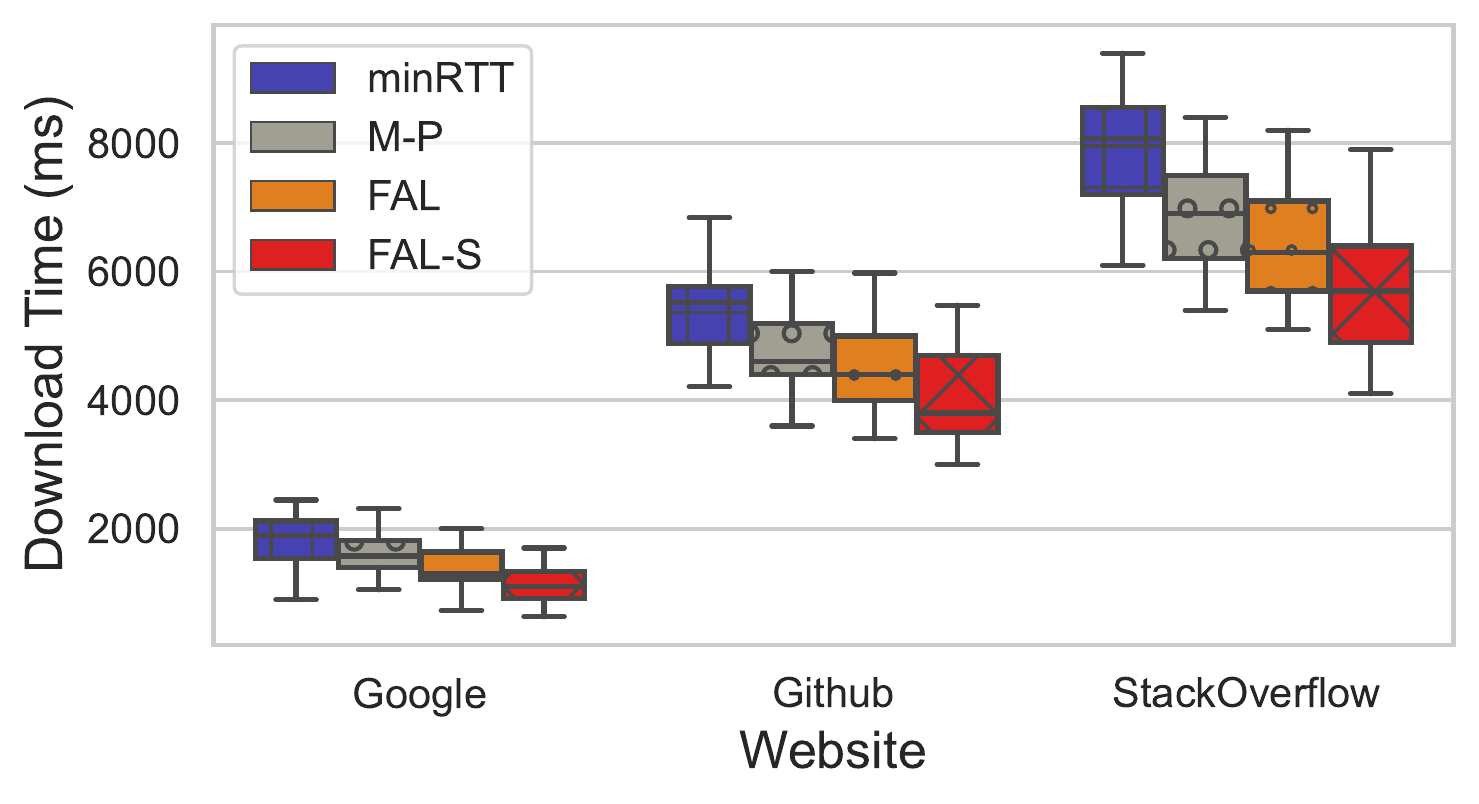}
	\caption{\rev{Web download test of FALCON, M-Peekaboo, and minRTT.}}
	\label{webpage}
\end{figure}

\subsubsection{Stress Test}
We also perform a stress test to examine how fast and accurately FALCON, DQN-On, and M-Peekaboo adapt to changing network conditions. In order to isolate the impact of adaptation, none of the schedulers has buffered online data beforehand.
We define a change interval and, under each change interval, we generate 24 different network conditions where the characteristics of each path are randomly generated in the range formed by the minimum and maximum of the characteristics shown in Table~\ref{tab_parameter}. 
At the end of each change interval, we calculate the relative score of the multipath scheduler using the approach presented in Section~\ref{emulate::dqnconv}.

Figure~\ref{fig_dqn stress} shows the relative score of each multipath scheduler under the stress test with change intervals of 8.0, 4.0, 2.0, 0.5, and 0.3 seconds, respectively. 
We observe that DQN-On already struggles when the change interval is 8.0 seconds, as indicated by a relative score much less than 1. This is in line with the results on the convergence behavior of DQN-On in Section~\ref{emulate::dqnconv}. We also observe that M-Peekaboo struggles with a change interval of 2 seconds, as indicated by the drop of its relative score compared to the scores obtained with change intervals of 8.0 and 4.0 seconds. We further observe that FALCON performs very well up to a change interval of 0.5 seconds. Then, it experiences a performance drop when the change interval is equal to 0.3 seconds. When both FALCON and M-Peekaboo can catch up with the change of network conditions (e.g., the change interval of 4 seconds), FALCON reaches a higher performance than M-Peekaboo. In all cases, FALCON shows higher scores compared to all the other schedulers, ultimately highlighting a significantly higher both adaptation accuracy and speed.

\subsection{Multi-streaming support for Web Services}
\label{emulate::extend}

\rev{
In Sections \ref{emulate::static} and \ref{emulate::closerlook}, we have examined the effectiveness of FALCON for bulk transfer services. In this section, we will examine the extensibility of FALCON for web services.} 

\rev{For the web experiments, we use the stream multiplexing feature of MPQUIC, an important feature that is planned to be exploited in HTTP/3. Therefore, we follow the approaches in the existing literature for dealing with stream multiplexing, and utilize a weighted round robin stream scheduling approach to download webpage objects based on their position in the dependency tree of the webpage ~\cite{wang2019multipath, rabitsch2018stream,langley2017quic}.
Moreover, considering the multi-streaming feature of MPQUIC, we plug in partially different contents within the framework of FALCON for web download from that for bulk transfer. Although both are subject to the same algorithm (i.e. single streaming is a special case of multi-streaming),  we denote the one with multi-streaming support as FALCON-S and FALCON for the one with single-streaming support, just to ease the presentation in the experiment. There are two main differences between FALCON and FALCON-S: 1) FALCON-S takes the send windows of each object stream as the state information while FALCON treats the send window as a whole for the state information; 2) FALCON-S splits the congestion window based on the weights of concurrent streams as the state information for each stream while FALCON treats the congestion window as a whole for the state information.}

\rev{ We perform the web experiment within the mobile scenario as defined in Section~\ref{emulate::static}, to better illustrate the algorithm's adaptation ability.
Figure~\ref{webpage} shows the download time of minRTT, M-Peekaboo, FALCON, and FALCON-S for different web pages. We observe that FALCON still has a clear performance gain over the other multipath schedulers. Furthermore, FALCON-S outperforms FALCON, reaching up to 13.6\% shorter download time. The results shows that it is possible to use FALCON across different applications, indicating the robustness of FALCON. Moreover, simple application-specific tuning can be applied to FALCON, in order to customize it for specific applications, eventually indicating the flexibility of FALCON.} 

\begin{figure}[t]
	\centering 
	\subfigure[\footnotesize Static: 5G and WLAN] {\includegraphics[draft=false,width=0.49\columnwidth]{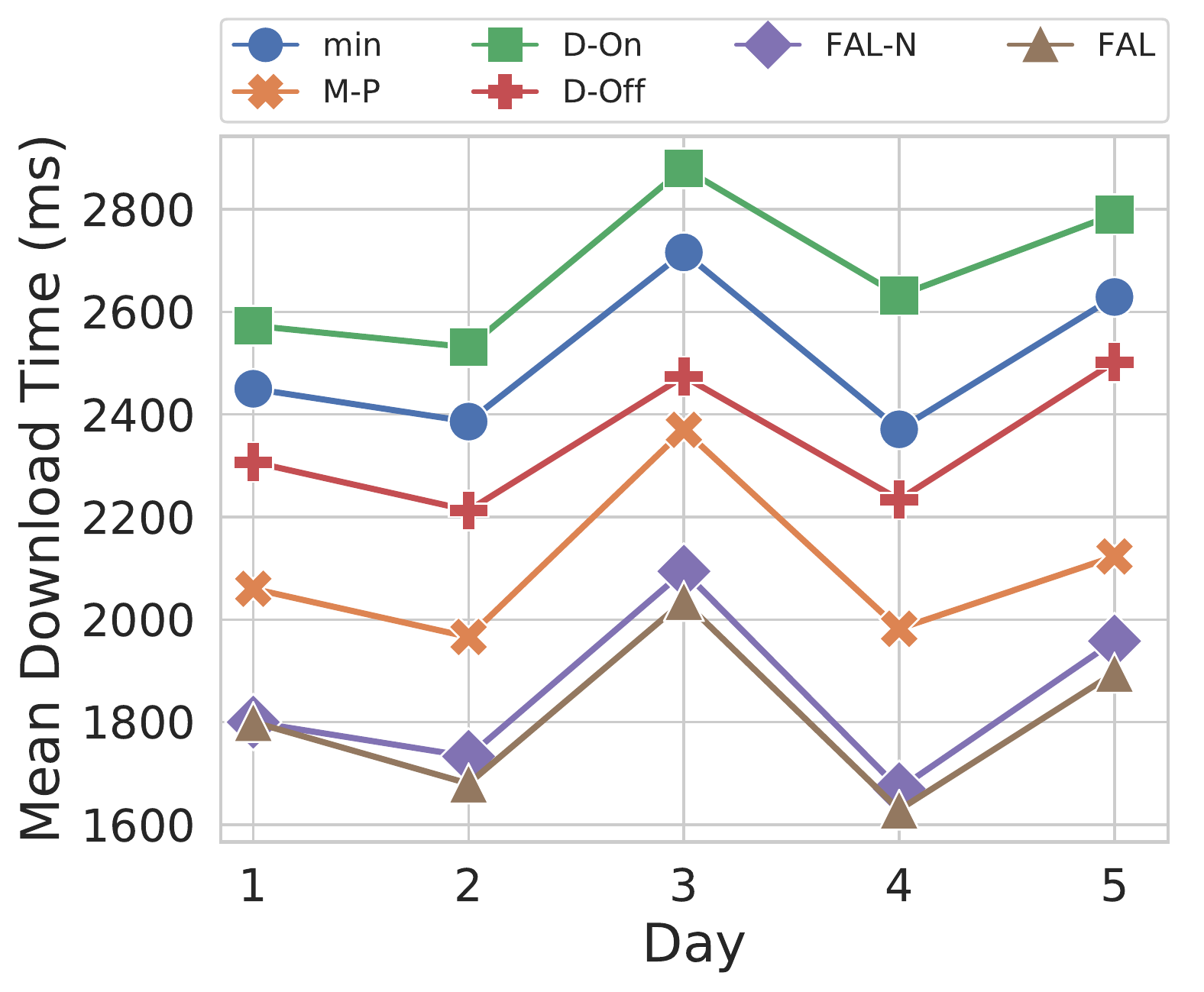}}
	\subfigure[\footnotesize Driving: 4G and 5G] {\includegraphics[draft=false,width=0.49\columnwidth]{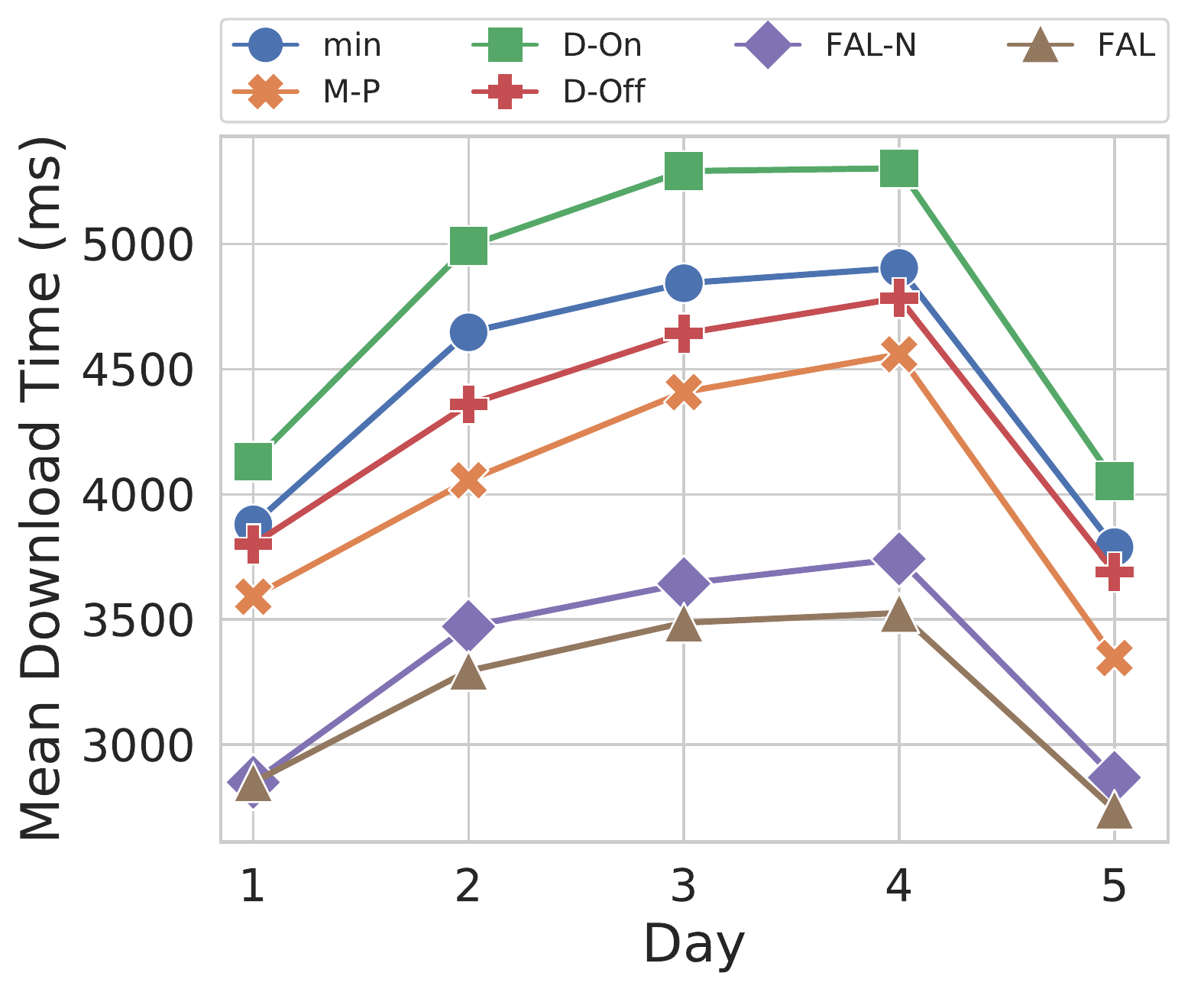}}
	\caption{Performance of FALCON and other multipath schedulers in real-world static and mobile network conditions over 4G, 5G, and WLAN.}
	\label{fig_r5gwlan}
\end{figure}

\section{Real-world Experiments}
\label{realworld}

We now present the evaluation of the schedulers in real-world experiments in both static and mobile scenarios. 

The static scenario is set up over 5G from a network provider and WLAN, while the mobile scenario is set up for a vehicle moving at a nearly constant speed of 30 km/h, over 5G from the same network provider and 4G from a different network provider. The evaluation is carried out in the afternoon, while the data for creating the meta-models in FALCON and for training DQN-Off is collected 5 days before the evaluation, in the morning. 
We perform the evaluation over 5 consecutive days and, at the end of each day, FALCON performs the offline update of the meta-models. To illustrate the effect of the offline update, we also show the performance of FALCON with no offline update, denoted FALCON-N (in these settings, FALCON and FALCON-N are identical in the first day of evaluation).

Figure~\ref{fig_r5gwlan} illustrates the performance of FALCON and other multipath schedulers in real-world network conditions. We note that in both static and mobile scenarios, the performance of DQN-On is always lower compared to all schedulers, due to the frequent retraining as the network conditions change. While in the emulated environment, DQN-On can converge (see Section~\ref{emulate::dqnconv}), in the real-world environment, even in the static scenario, the state transitions are more frequent due to the dynamicity of real networks. DQN-Off performs consistently better than DQN-On and minRTT, but worse than M-Peekaboo, FALCON, and FALCON-N, as it lacks the ability to adapt online. Restricted by its adaptation time, M-Peekaboo has worse performance in the mobile scenario than in the static case. FALCON and FALCON-N outperform M-Peekaboo with up to 23.6\% and 18.7\% shorter mean download time, respectively. In particular, FALCON outperforms FALCON-N, with a gain indicating that the effect of updating the meta-models is incremental. 
To summarize, the results show a rather high 
generalizability of the meta-models learned over the distribution of network conditions.

\section{FALCON's Configuration Parameters and Overhead} 
\label{indepth}
In this section, we study the impact of FALCON's configuration on the obtained performance  (Section~\ref{indepth::study}). We then discuss the overhead of FALCON (Section~\ref{indepth::overhead}). 

\subsection{A Study into FALCON's Configuration}
\label{indepth::study}

\rev{We study the impact of the configuration parameters adopted for FALCON on the observed performance, e.g. adaptation speed and accuracy. More  specifically, we address the selection of $K$ and the number of meta-models, which are directly related to FALCON operations.}

\subsubsection{Selection of $K$}
\label{indepth::selectionK}
\rev{
We first study the selection of $K$ and its impact on adaptation speed and accuracy. Recall that $K$ is the number of online training steps to be performed for fine-tuning the pre-built meta-models. We expect that for any given network condition, the online model should converge within $K$ steps. Since we perform a mini-batch gradient descent with the size of 32 for each step, the learning overhead in terms of the number of packets becomes the number of steps multiplied by the mini-batch size. Thus, we seek for the smallest value of $K$ that guarantees fast and accurate adaptation.}

\begin{figure}[t]
	\centering 
\includegraphics[draft=false,width=0.88\columnwidth]{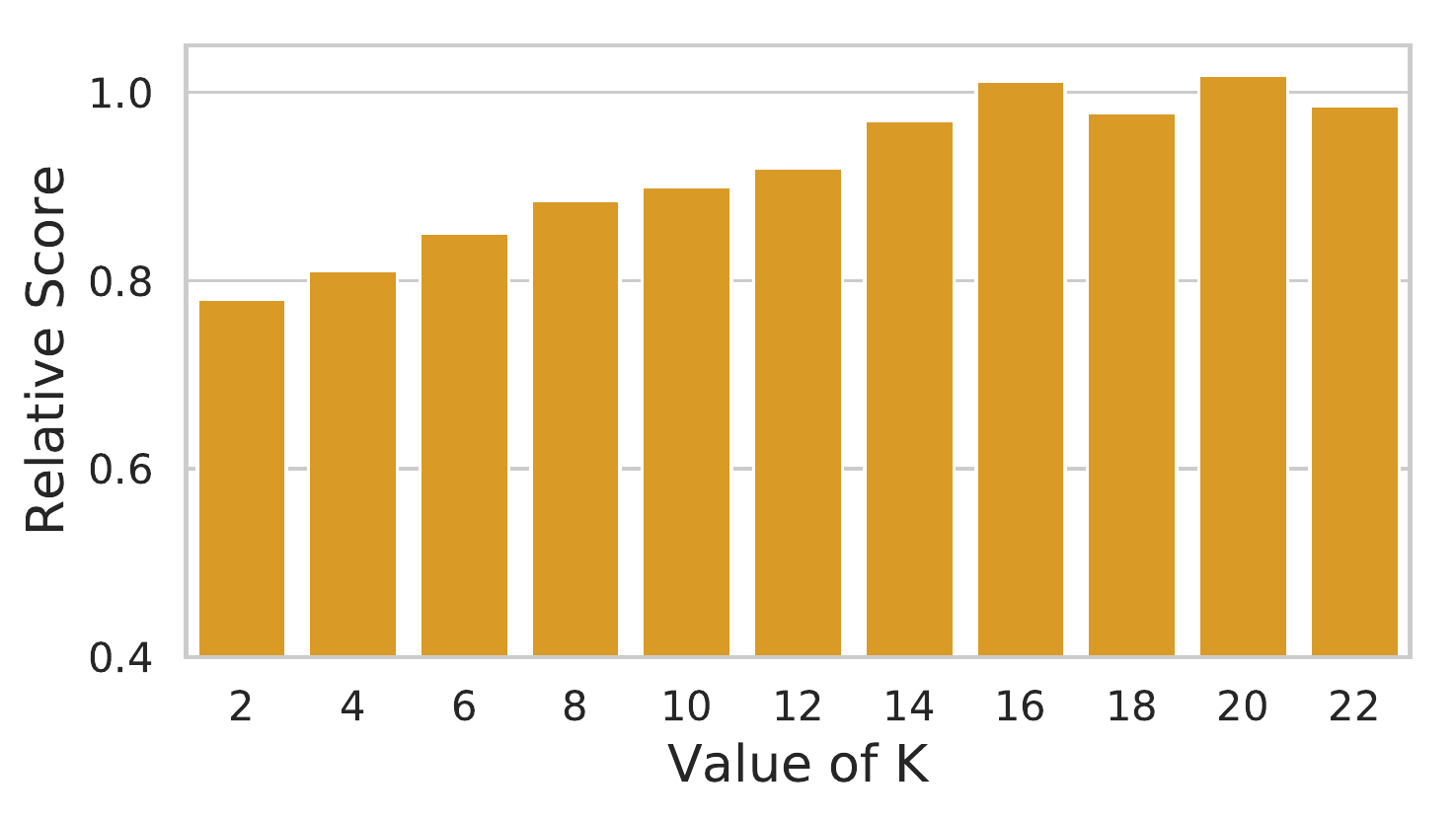}
	\caption{\rev{The impact of K on FALCON's performance.}}
	\label{fig_dqnK}
\end{figure}

\rev{
Figure~\ref{fig_dqnK} shows the performance of FALCON in terms of relative score as a function of $K$. We first observe that, when $K$ is relatively small, FALCON does not show significant gains. This is because the meta-learning mechanism is struggling to find meta-models that can converge within the $K$ steps. 
The performance saturates when $K=16$, which is therefore selected as the parameter adopted in FALCON. }

\begin{figure}[t]
	\centering 
\includegraphics[draft=false,width=0.88\columnwidth]{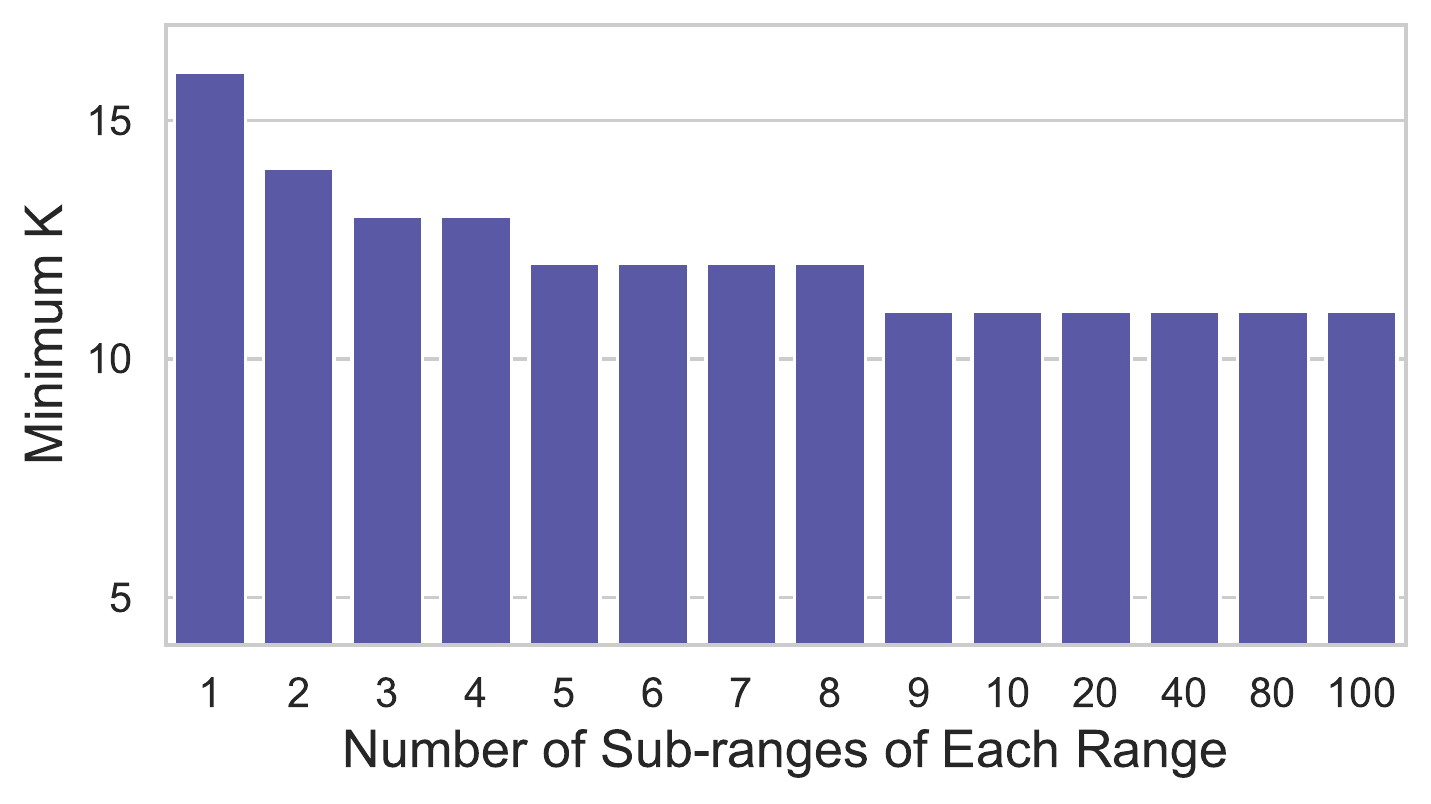}
	\caption{\rev{The impact of the number of sub-ranges (also, the number of meta-models) on the optimal K.}}
	\label{fig_dqnnum}
\end{figure}

\subsubsection{Number of Meta-models}
\label{indepth::number}
\rev{
Next, we study the impact of the number of meta-models we employ. Recall that, for a combination of our defined range of link characteristics, we train one meta-model to bootstrap. To obtain a higher amount of meta-models over the defined range of link characteristics, we divide each range into multiple sub-ranges (e.g., divide the $[0, 1)$\% range of loss rate into a number of sub-ranges) and train one meta-model for each combination of sub-ranges.}

\rev{
Figure~\ref{fig_dqnnum} shows the minimum value of $K$ (as analyzed  in Section~\ref{indepth::selectionK}) as a function of the number of sub-ranges for each range of link characteristics (the original one is 1). We observe that the value of $K$ slowly decreases as we increase the number of sub-ranges for each range. However, even if the number is set to a relatively large number (e.g., 100), the minimum $K$ is still relatively large. Indeed, the meta-model still requires a certain number of training steps before converging to the optimal values.} 

In theory, when the number of sub-ranges (also, the number of meta-models) is sufficiently high, i.e., offline and online scenarios will converge, the minimum value of $K$ will be zero, meaning that there will be no need for online adaptation. Nevertheless, this is not practical for the reasons we present in Section~\ref{problem_statement}. Furthermore, FALCON requires the estimation of current network conditions, in order to map such conditions to one of the pre-built meta-models. 
The estimation error can easily cause disturbance in selecting the meta-models if the number of meta-models is too large. 
For this reason, we keep the original number of meta-models for FALCON that is practical and avoids the estimation errors with the satisfactory performance and adaptation speed.  

\begin{figure*}[t]
	\centering 
	\subfigure[\footnotesize CPU Cost] {\includegraphics[draft=false,width=0.88\columnwidth]{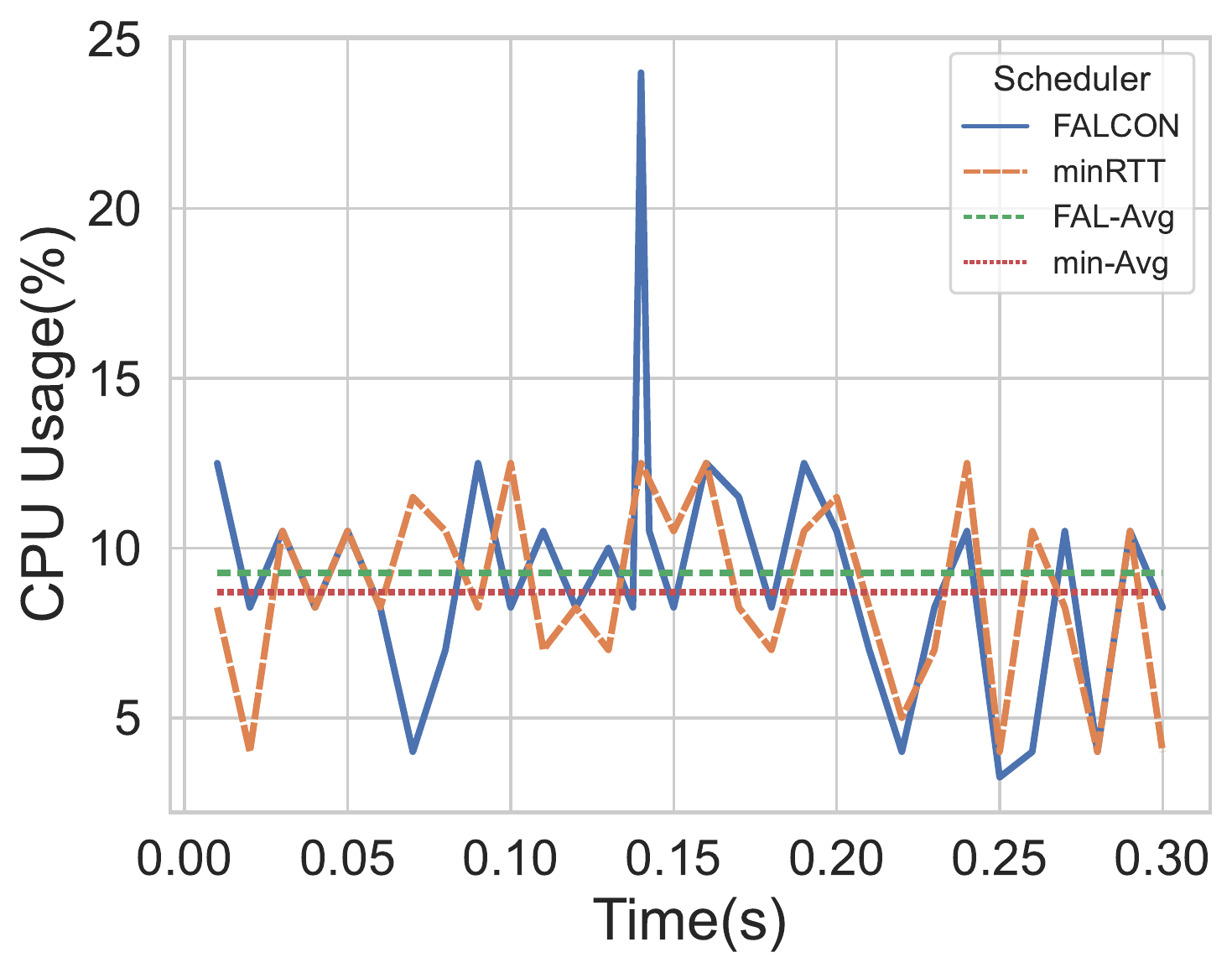}}
	\subfigure[\footnotesize Memory Cost] {\includegraphics[draft=false,width=0.88\columnwidth]{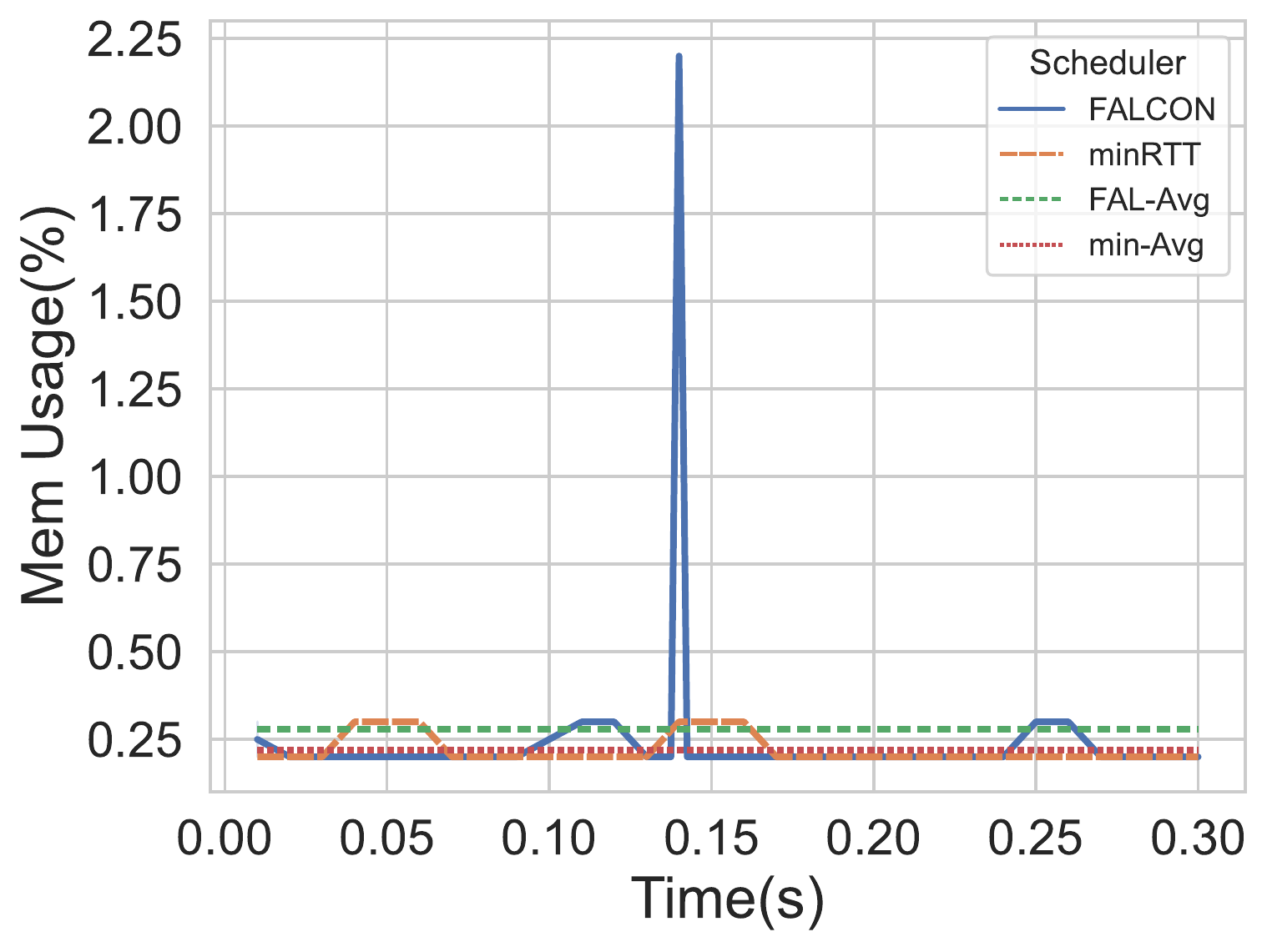}}
	\caption{\rev{Overhead of FALCON and minRTT in the perspectives of CPU usage percentage and memory usage percentage.}}
	\label{fig_overhead}
\end{figure*}

\subsubsection{Discussion on Hyperparameter Selection}
\label{indepth::hyper}

Learning based systems cannot avoid the necessity to employ hyperparameters in their algorithms. The search for the hyperparameters to use is an optimization problem, which is often solved heuristically in a trial-and-error manner. 
In extreme cases, the trial-and-error process can be automated, and this is known as automated machine learning~\cite{yao2018taking}. In all cases, the higher the complexity of the model and the task, the longer the time per trial would be. Thus, this approach is normally used with small-sized models and datasets so that the iterations for  optimization can be completed until a set of parameters is found. 
Since FALCON is set up over a significant size of models and data, this optimization approach is not feasible for FALCON, just as for most of the other practical machine learning systems that eventually employ intuitive hyperparameters with human-in-the-loop manual optimization (tuning). Therefore, as regards to DQN-related parameters, we adopt the common parameters, since they already bring significant gains for FALCON against other schedulers. We further observe that the selection of these parameters is subject to machine learning engineering aspects, and their optimization may result in further improvements.

\subsection{System Overhead of FALCON}
\label{indepth::overhead}

A deep learning system normally consists of a training phase and an inference phase (i.e., the interpretation of the neural network). For FALCON, the training phase is partially completed offline and partially completed online along with the inference phase; the inference phase is completed online. The offline training phase is of higher system overhead but free of impact on the deployment of FALCON, since it occurs in an offline manner.

We perform an experimental analysis to investigate the system overhead of FALCON in the online phase. We record the central processing unit (CPU) usage and memory usage of FALCON and minRTT over a 0.3 seconds period with a single change of network conditions. The CPU of the server is Intel-i5 (2.50 GHz) of two cores and the size of the memory of the server is 8 GB. Figure~\ref{fig_overhead} shows both the real time usage and the average usage of FALCON and minRTT. 
First, we observe that the average CPU usage of FALCON is only 3\% higher than minRTT and the memory usage of FALCON is on average only 6\% higher than minRTT.  
Recalling that we have a change of network conditions within the analyzed time frame of 0.3 seconds, i.e., the worst-case scenario where FALCON can be used,
the CPU and memory usage gaps between FALCON and minRTT are likely to be even smaller when the network conditions change less frequently. Thus, overall FALCON does not bring significant system overhead. 
For the real time usage, we do observe a significant spike in terms of CPU and memory usage. 
This spike happens when FALCON performs the online training with the gradient calculation. 
However, we do not observe any extra system overhead in the online inference phase, because FALCON utilizes a neural network model with a relatively simple architecture and thus with low computational complexity. 
Lastly, FALCON is deployed in the server in the context of this paper, thus the extra CPU and memory costs are not as significant issues as they would be on the client device.

However, we do not limit the applicable scenarios of FALCON to server side deployment. A path selection on the client side can also employ FALCON. In such a context, the client would be a mobile device, constrained by power consumption. For the online inference side, as mentioned above, FALCON and other schedulers should hold similar power consumption, as inferred from the similar CPU and memory utilization. Considering a scenario where FALCON might employ a Neural Network (NN) model of higher complexity, and thus larger inference overhead, embedded software and hardware solutions such as ARM Common Microcontroller Software Interface Standard (CMSIS) NN software library~\cite{arm2021}, Field-Programmable Gate Array (FPGA), and Graphics Processing Unit (GPU), can make the inference more efficient.

\section{Discussion and Conclusion}
\label{conclusion}

Learning-based networking systems have received much attention of late, as well intriguing the field of multipath scheduling. However, the deployment of existing learning-based multipath schedulers fails to be functional in the aspects of achieving a fast and accurate adaptation. 

In this paper, we propose FALCON, a learning-based multipath scheduler that can adapt fast and accurately to time-varying network conditions by combining the benefits of online and offline learning. 
Through extensive emulations, we show that FALCON is able to consistently outperform all state-of-the-art schedulers by adapting to the network conditions in a fast and accurate manner. Our real-world experiments confirm that FALCON performs well also under realistic network settings. 

We see two main future directions for this work. Firstly, in this paper, we have demonstrated the possibility of applying DQN within FALCON, but we will also consider applying other deep learning approaches to enhance the performance of FALCON.
Secondly, we plan to interpret and understand the learning outcome of FALCON (i.e., in the form of NN) to potentially deduce the guaranteed performance bound.

\bibliographystyle{IEEEtran}
\bibliography{sample-base}

\newpage

\begin{IEEEbiography}[{\includegraphics[width=1in,height=1.25in,clip,keepaspectratio]{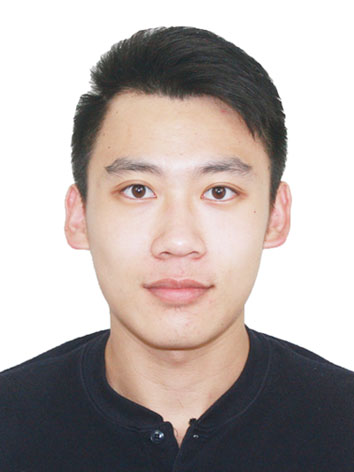}}]
	{Hongjia Wu}
	is a Ph.D. candidate at Simula and OsloMet. He obtained his M.Sc. in Embedded Systems from TU Delft and B.Sc in Automatic Control from Northeastern University. His research interests include multipath protocols and robotic systems. 
\end{IEEEbiography}

\vskip 0pt plus -1fil
\vspace{-10mm}

\begin{IEEEbiography}[{\includegraphics[width=1in,height=1.25in,clip,keepaspectratio]{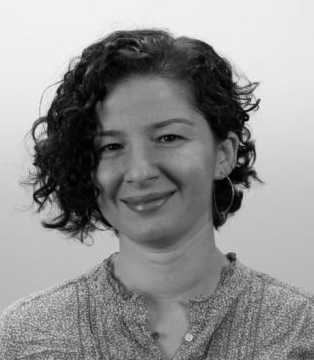}}]
	{\"{O}zg\"{u} Alay}
	Dr. Ozgu Alay received the B.S. and M.S. degrees in Electrical and Electronic Engineering from Middle East Technical University, Turkey, and Ph.D. degree in Electrical and Computer Engineering at Tandon School of Engineering at New York University. Currently, she is an Associate Professor in University of Oslo, Norway and Head of Department at Mobile Systems and Analytics (MOSAIC) of Simula Metropolitan, Norway. Her research interests lie in the areas of mobile broadband networks, multipath protocols and robust multimedia transmission over wireless networks. She is author of more than 70 peer-reviewed IEEE and ACM publications and she actively serves on technical boards of major conferences and journals.
\end{IEEEbiography}

\vskip 0pt plus -1fil
\vspace{-10mm}

\begin{IEEEbiography}[{\includegraphics[width=1in,height=1.25in,clip,keepaspectratio]{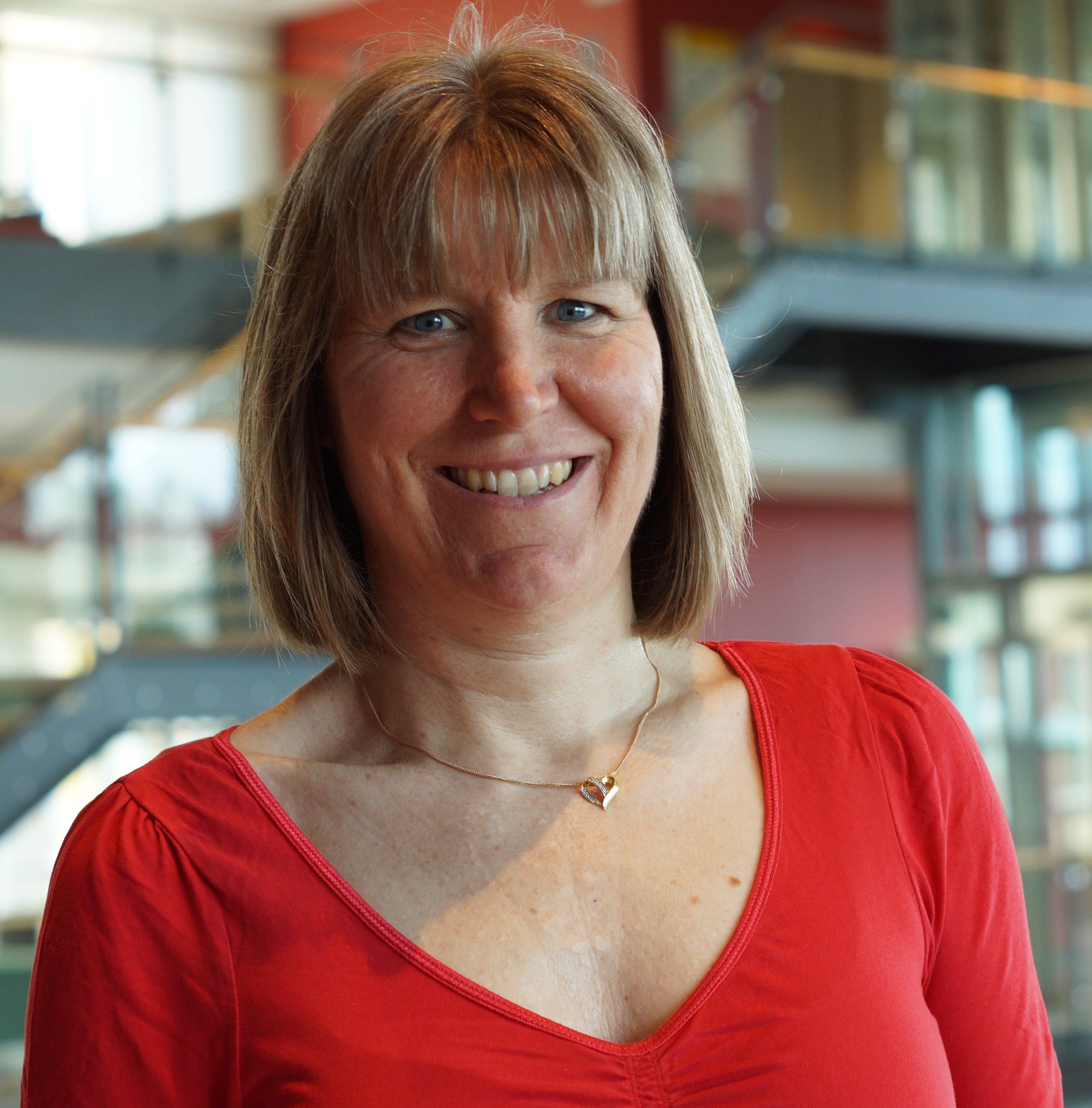}}]
	{Anna Brunstrom}
	received a B.Sc. in Computer Science and Mathematics from Pepperdine University, CA, in 1991, and a M.Sc. and Ph.D. in Computer Science from College of William \& Mary, VA, in 1993 and 1996, respectively. She joined the Department of Computer Science at Karlstad University, Sweden, in 1996, where she is currently a Full Professor and Research Manager for the Distributed Systems and Communications Research Group. Her research interests include Internet architectures and protocols, techniques for low latency Internet communication, multi-path communication and performance evaluation of mobile broadband systems including 5G. She has authored/coauthored over 170 international peer-reviewed journal and conference papers.
\end{IEEEbiography}

\vskip 0pt plus -1fil
\vspace{-10mm}

\begin{IEEEbiography}[{\includegraphics[width=1in,height=1.25in,clip,keepaspectratio]{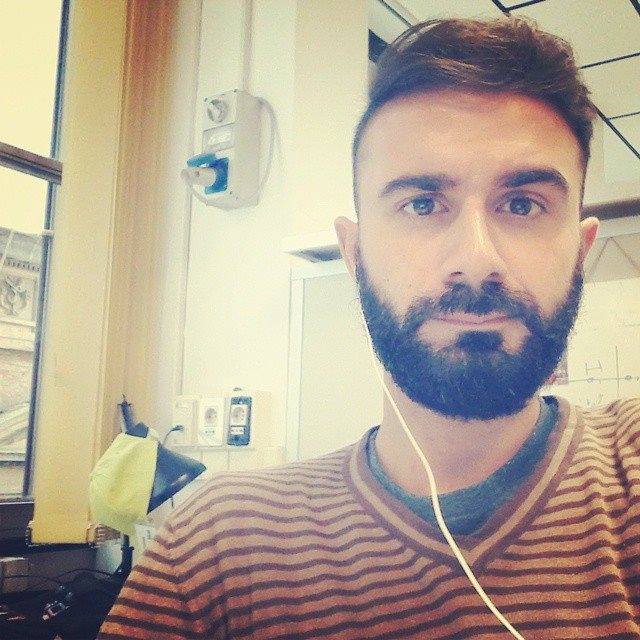}}]
	{Giuseppe Caso}
	is an Experienced Researcher at Ericsson Research (Radio Systems and Standards) in Kista, Sweden. In 2018-2021, he was a Postdoctoral Fellow with the MOSAIC Department at SimulaMet, Oslo, Norway. In 2016, he received the Ph.D. degree from Sapienza University of Rome, where he was a Postdoctoral Fellow until 2018. From 2012 to 2018, he has held visiting positions at Leibniz University of Hannover, King's College London, Technical University of Berlin, and Karlstad University. His research interests include cognitive and distributed communications,  resource allocation in cellular systems, IoT technology and evolution, and location-based services. He is an IEEE Member.
\end{IEEEbiography}

\vskip 0pt plus -1fil
\vspace{-10mm}

\begin{IEEEbiography}[{\includegraphics[width=1in,height=1.25in,clip,keepaspectratio]{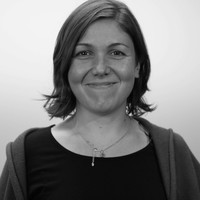}}]
	{Simone Ferlin}
	is a software researcher at Ericsson AB in radio networks. She received her Dipl.-Ing. degree in Information Technology with major in Telecommunications from Friedrich-Alexander Erlangen-Nuernberg University, Germany in 2010 and her PhD degree in computer science from the University of Oslo, Norway in 2017. Her interests lie in the intersection of cellular networks and the Internet, with her research focusing on computer networking, QoS and cross-layer design, transport protocols, congestion control, network performance, security, and measurements. Her dissertation focused on improving robustness in multipath transport for heterogeneous networks with MPTCP. She actively serves on technical boards of major conferences and journals in these areas. 
\end{IEEEbiography}

\end{document}